\documentclass[twocolumn,aps,pra,superscriptaddress,amsmath,amssym,floatfix,longbibliography,nofootinbib]{revtex4-1}
\usepackage{lipsum}
\usepackage{graphicx}
\usepackage{bm}
\usepackage{dsfont} 
\usepackage{amsmath}
\usepackage{color}
\usepackage[usenames,dvipsnames]{xcolor}
\usepackage{tikz}
\usetikzlibrary{%
    decorations.pathreplacing,%
    decorations.pathmorphing,%
    arrows
}
\usepackage{amssymb}
\usepackage{enumerate}
\usepackage{comment}
\usepackage{listings}
 \usepackage{latexsym} 
\usepackage{revsymb} 
\usepackage{accents}

\usepackage[colorlinks=true,linkcolor=MidnightBlue,citecolor=blue,filecolor=green,urlcolor=PineGreen]{hyperref}

\def\la{\langle}
\def\ra{\rangle}

\def\be{\begin{equation}}
\def\ee{\end{equation}}

\newcommand{\ket}[1]{\left\vert #1 \right\rangle}
\newcommand{\bra}[1]{\left\langle #1 \right\vert}

\newcommand{\idop}{\mathds{1}}

\definecolor{codegreen}{rgb}{0,0.6,0}
\definecolor{codegray}{rgb}{0.5,0.5,0.5}
\definecolor{codepurple}{rgb}{0.58,0,0.82}
\definecolor{backcolour}{rgb}{0.95,0.95,0.95}
\definecolor{internationalorange}{rgb}{1.0, 0.31, 0.0}
\definecolor{cadetgrey}{rgb}{0.57, 0.64, 0.69}

\lstdefinestyle{mystyle}{
    backgroundcolor=\color{backcolour},   
    commentstyle=\tt\color{Cerulean},
    keywordstyle=\tt\color{Fuchsia},
    numberstyle=\it\tiny\color{MidnightBlue},
    stringstyle=\tt\color{red},
    basicstyle=\tt\footnotesize,
    breakatwhitespace=false,         
    breaklines=true,                 
    captionpos=b,                    
    keepspaces=true,                 
    numbers=left,                    
    numbersep=5pt,                  
    showspaces=false,                
    showstringspaces=false,
    showtabs=false,                  
    tabsize=2
}
 
\begin{document}

\title{Quantum measurement engines and their relevance for quantum interpretations}
\author{Andrew N. Jordan} 
\affiliation{Department of Physics and Astronomy, University of Rochester, Rochester, NY 14627, USA}
\affiliation{Institute for Quantum Studies, Chapman University, Orange, CA 92866, USA}
\author{Cyril Elouard}
\affiliation{Department of Physics and Astronomy, University of Rochester, Rochester, NY 14627, USA}
\author{Alexia Auff\`eves}
\affiliation{Institut N\'eel, CNRS/UGA
25, avenue des Martyrs - BP 166
Fr-38042 Grenoble Cedex 9, France}
\date{\today}

\begin{abstract}
   This article presents recent progress in the theory of quantum measurement engines and discusses the implications of them for quantum interpretations and philosophical implications of the theory.  
Several new measurement engine designs are introduced and analyzed: We discuss a feedback based atom-and-piston engine that sharply associates all work with successful events and all quantum heat with the failed events, as well as an unconditional but coherent qubit engine that can attain perfect efficiency.  
Any quantum measurement of an observable that does not commute with the Hamiltonian will necessarily change the energy of the system. We discuss different ways to extract that energy, the efficiency and work production of that process.  
\end{abstract}

\maketitle

\section{Introduction} \label{sec:intro}
The ``Quantum Limits of Knowledge'' Copenhagen conference brought together many experts of the foundational aspects of quantum physics back to the birthplace of the still-dominate interpretation of quantum physics.  Despite its assault in the philosophical and physical literature, the Copenhagen interpretation remains the default rough and ready interpretation of quantum mechanics, deriving from the thinking of Niels Bohr \cite{Bohr28}. Quantum physics - as a science - provides a set of testable predictions a formalism and set of prescriptions to apply to carry out a comparison of theory and experiment.  Quantum physics needs no more than this.  Strictly speaking from a scientific point of view, no further interpretation is necessary, as is evident from the history of interpretational disagreements providing no barrier to quantum physics becoming our most accurate theory of nature.

Nevertheless, as scientists interested in philosophical matters, we may wish for more.  There can be many reasons for this.  On one hand, we may wish to have a deeper understanding of the meaning of world, and to the extent that science can provide any insight into this matter, it is certainly worth pursuing.  As working scientists, it may be that quantum physics is not the last word on physical description of the microscopic world, and that applying pressure on the weak points of the theory will point us to a deeper theory of nature.  Finally, even though a developed metaphysics of quantum mechanics is not necessary to use the theory, we often have a working philosophical picture, which may guide (or misguide) us in the quest to discover new effects and phenomena within quantum mechanics.

\begin{figure*}[htb]
\centering
\includegraphics[scale=0.57]{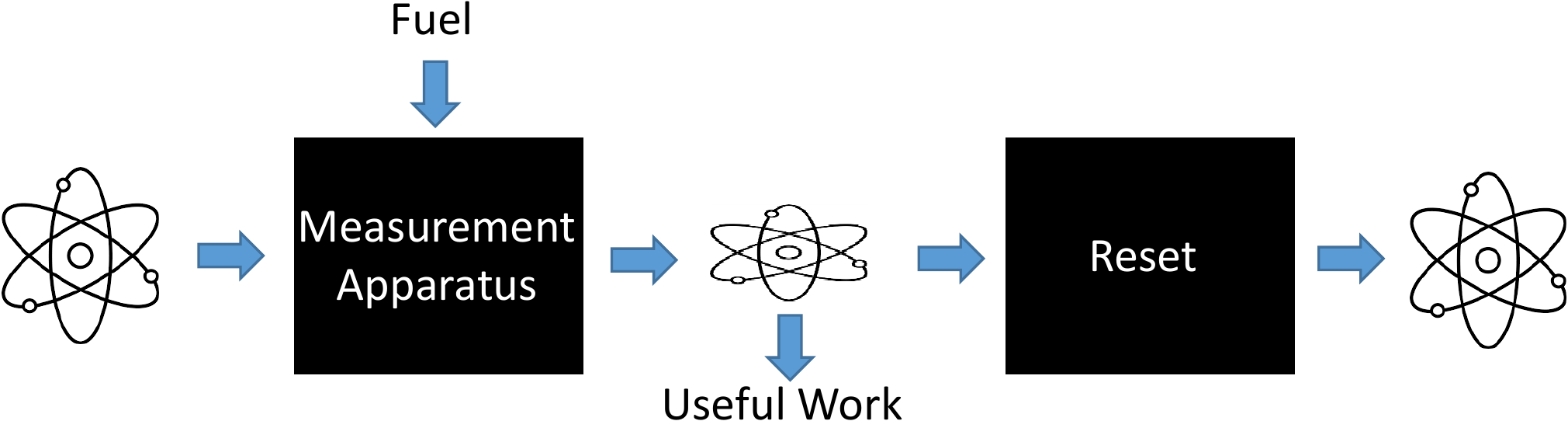}
\caption{Schematic of the operating of the quantum measurement engine.  A quantum system is measured via an ancillary system (measurement apparatus). Energy may be given to the quantum system in the process, which is then extracted as useful work.  The system is reset via coherent or incoherent processes, resetting the system for the next cycle.  For repeated cycles to work, the measurement apparatus must be re-energized.}
\label{fig:localwave}
\end{figure*}

The subject of this article, quantum measurement engines, is the application of quantum mechanical principles to creating systems capable of doing useful work at the quantum level.
The basic idea being discussed is illustrated in Fig.~1.  We wish to do useful work on a single quantum system.  The work is carried out by measuring an observable of the quantum system that does not commute with the Hamiltonian of the system alone.  The measurement needs not be projective on the system - it can refer to any process where an auxillary body is first entangled with the system of interest, followed by measurement of that body. Such a measurement accomplishes two important things.  First, it gives a result (or outcome) that allows an inference about the system of interest. Second, it generally results in a change of the quantum state of the system, usually called wavefunction collapse \cite{WisemanBook}. Such collapse processes, both complete, partial, continuous, and weak have been extensively investigated experimentally in the last decades, with excellent agreement with theoretical descriptions (see e.g. Refs.~\cite{Nogues99,korotkov2006undoing,katz2008reversal,williams2008weak,dixon2009ultrasensitive,goggin2011violation,Kocsis11,chantasri2013action,Murch13,jordan2013quantum,weber2014mapping,viza2015experimentally,chantasri2016quantum,naghiloo2017quantum,chantasri2018simultaneous,Dassonneville19,Rossi19}). With the change of state, the disturbance of the system also implies a change of expectation values of observables.  We are particularly concerned here with the Hamiltonian of the system, but could also consider observables such as linear and angular momentum, spin components, composite observables, and so on.  The generalized measurement permits us to know what the disturbance of the quantum system is, as well as know what conditional energy is given or taken away from it.  This knowledge in turn permits us to take a feedback action such as altering the potential of the system or applying some unitary operation in order to extract the energy from the system.  In the sections below we will see that sometimes these actions are conditional, and sometimes they can be blind to the measurement outcomes.  The main point is that the energy can be extracted in the form of useful work, and following a reset operation (that may be dissipative in general), c.f. Fig.~1, an engine cycle is created. 

There is an analogy here to thermodynamic heat engines working between hot and cold reservoirs.  In that case, thermal energy in the form of heat is extracted from the hot reservoir, a portion of it is converted into useful work, and the remaining part of the heat is discarded into the cold reservoir.  Importantly, the entropy must be nonnegative on average according to the second law \cite{Clausius1879}, which bounds the efficiency of the engine to Carnot efficiency \cite{Carnot1824}.  The source of the energy in this case is a disordered energy reservoir characterized with a temperature.  In contrast, in the quantum measurement case, the energy source is the measurement device itself. This energy transfer is also intrinsically random, as it is associated with the random selection of the measurement outcome, which lead to the label ``quantum heat'' \cite{Elouard17Role}. However, in contrast with the heat provided by an actual thermal reservoir this energy may be at least partially ordered, and a measuring apparatus yields behavior somewhere in between a thermal reservoir and a work source.

More properly, it is a nonequilibrium energy resource, created for example, with resources of that type, using filters or conditional operations.  The important physical consequence of using this nonequilibrium state of the measuring device is the fact that there is no well-defined temperature of the energy source, and consequently no Carnot efficiency.  This means that for a well-chosen measurement process, one can in principle achieve perfect efficiency, that is, all energy contained in the meter degree of freedom can be perfectly converted into system work \cite{Elouard17}.

\section{Feedback-based measurement engines} \label{sec:theory}

The detailed understanding of the different ways the energy conversion can be accomplished is best seen with a few simple examples that have been proposed so far.  In this section, we focus on engines that exploit the measurement outcome to extract work via some feedback mechanism. This principle is reminiscent of Maxwell's demon type of engines, with the crucial different that the latter use measurement to extract the energy from a hot thermal reservoir, while in the engines presented in this section, the measurement itself is the source of the energy and no hot thermal bath is involved.
\begin{figure}[bht]
\includegraphics[scale=.8]{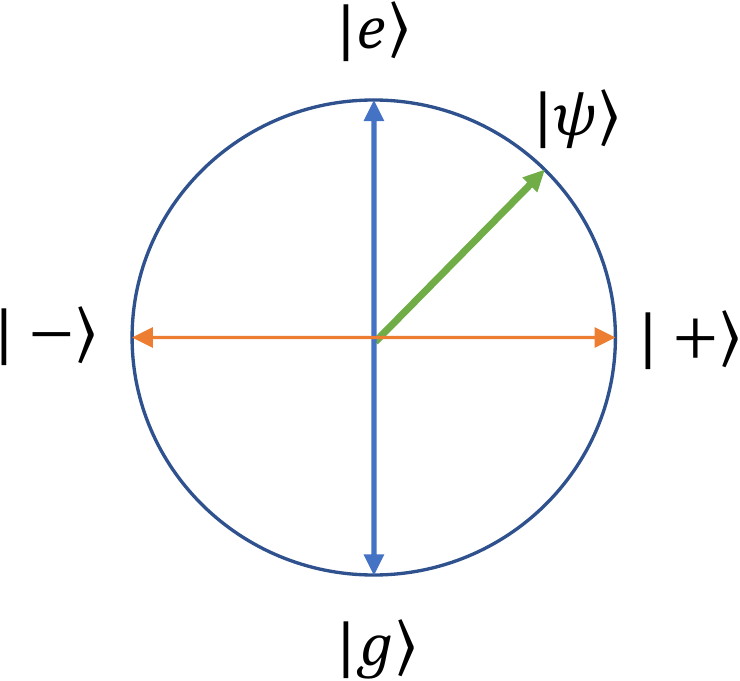}
\caption{A great circle of the Bloch sphere is illustrated.  The quantum state (green arrow) dictates the probability to be found in the excited/ground state (blue double-sided arrow), or in the $|+\ra/|-\ra$ state (orange double-sided arrow).  The projection of the state $|g\ra$ into the $|\pm\ra$ basis always raises the energy of the system.}
\label{fig:bloch}
\end{figure}

\subsection{Qubit engine}
\label{sec:qubit}
The easiest measurement based engine to understand corresponds to Ref.~\cite{Elouard17} and involves a simple qubit that is being driven by a Rabi tone from electromagnetic radiation.  We have the choice to measure the qubit in whatever basis we like.  Let the ground and excited state of the qubit be $|e\ra, |g\ra$.  Suppose we prepare the qubit in the state $|\psi\ra = \alpha |g\ra + \beta |e\ra$.  The Hamiltonian is given by the operator
\be
H = \frac{\epsilon}{2} (|e\ra \la e| - |g\ra \la g|),
\ee
where $\epsilon$ is the energy splitting of the qubit.  It is clear that the expected energy of the system is $\la \psi | H |\psi \ra = (\epsilon/2) (|\beta|^2 - |\alpha|^2)$.

A measurement of the system in the ground/excited energy basis will result in either the state excited (with energy $\epsilon/2)$ so the system gains energy, or the state ground (with energy $-\epsilon/2$), so the system loses energy.  The observant reader will quickly notice that if we average the energy gain over the probability of the outcomes ($|\alpha|^2$ and $|\beta|^2$, respectively), with identically prepared initial conditions, it is clear that no gain or loss of energy can happen on average.  This is because we are measuring in a basis in which the Hamiltonian is diagonal, and it is clear that this feature generalizes to more complicated systems.

On the other hand, suppose that we measure in the basis $|+\ra, |-\ra$, where these states are defined as $|\pm \ra = (|e\ra \pm |g\ra)/\sqrt{2}$.
The measurement must return one of these values, and both of them have an expected value of 0, so the energy of the post-measurement state is always 0, which unless the initial state is on the equator of the Bloch sphere, will change the energy of the system.  This effect is particularly dramatic if we start with the ground state.  Then, the system will always gain energy $\epsilon/2$ following a measurement!  The now energized quantum system may be viewed as an energy resource.  We can extract the energy of the system by applying a $\pi/2$ pulse with an electromagnetic signal, that will take the energy from qubit and give it to the electromagnetic pulse.  There is a subtlety here - we must know what phase to give to the electromagnetic pulse so as to take the energy away (rather than give it).  To know this, we must pay attention to the result from the first measurement.  If we obtain the $|+\ra$ state, we apply a $\pi/2$ pulse, whereas if we obtain the $|-\ra$ state, we apply a $-\pi/2$ pulse.  The measurement plus the feedback sequence then closes an engine cycle which may be repeated indefinitely.  

It is a natural question to ask where the energy comes from for this cycle.  The answer is that the system carrying out the measurement must be included in the analysis, which is typically removed in the discussion of making the quantum measurement.  Other works have shown that the system making the measurement must carry and deposit the energy to the system.  Repeated measurements must consequently have a reenergized meter, otherwise the engine will stall, having run out of fuel.  We can see this more explicitly by writing the total Hamiltonian as  $H = H_S + H_I + H_M$, so that the total system is described via a Hamiltonian approach.  The global unitary $U$ operating on both system and meter will then preserve the total expected energy, so loss of energy in one part of the whole must be made up in another part.

If one wishes to evaluate the efficiency of the engine, one must take into account the fact the memory of the measuring apparatus has to reset in order to fully close the engine cycle. This operation is required for any engine involving a measurement and feeback process, and was found essential to reconcile the Maxwell demon gedanken experiment with the second law of thermodynamics.  Works by Landauer, Bennett and others have lead to the conclusion that for a reversible memory reset, this operation was requiring a miminum work cost of $W_\text{reset} = k_\text{B}T S_\text{M}$, where $S_\text{M}$ is the entropy of the memory at the end of the cycle (namely, the Shannon entropy of the distribution of possible outcomes) and $T$ is the temperature of its surrounding bath \cite{landauer1961irreversibility,bennett2003notes}. Finally the efficiency of the conversion of the ``quantum heat'' ${\cal E}_\text{M} >0$ received by the system during the measurement, into work extracted $W \leq 0$ is $\eta = -(W + W_\text{reset})/{\cal E}_\text{M}$. In the ideal engine cycle we are interested in this section, all the energy coming from the measurement process is converted into work $W = - {\cal E}_\text{M}$, such that the efficiency solely differs from unity due to the erasure cost: $\eta = 1-W_\text{reset}/{\cal E}_\text{M}$.
 
This very simple engine provides a good illustration of the difference between the measurement process and a measurement channel. For this we consider that now after a measurement finding the qubit, say in state $\ket{+}$, only a very small fraction of the available energy is extracted, by performing a rotation of angle $\theta\ll 1$; i.e. $W = -\epsilon\sin(\theta)/2$. After such extraction, the state $\ket{\psi_\theta}$ of the qubit has a very large overlap with the state $\ket{+}$, verifying $\vert\bra{+}\psi_\theta\rangle\vert^2 = \cos^2(\theta/2)$. This implies that with very high probability $p_+ = \cos^2(\theta/2) \simeq 1$, another measurement will prepare the state $\ket{+}$ again. This high probability has a surprising consequences: First, the entropy of the memory vanishes, with a scaling $S_\text{M} \sim \theta^2 \log(\theta)$ when $\theta \to 0$, which is faster than the energy extracted from the measurement $W = - {\cal E}_\text{M} \sim \theta$.
Therefore, the efficiency reaches unity in the limit $\theta \to 1$. Second, the time needed to extract the work also scales like $\theta$, meaning that in the limit $\theta\to 0$ the power of the engine reaches a finite value \cite{Elouard17}. This result -- perfect efficiency at finite power -- witnesses that in this limit the measurement is a perfectly ordered energy source. Intuitively, one can see that for a very small angle $\theta$, the result $\ket{+}$ and the associated effect on the qubit become deterministic -- one need not read the outcome anymore to know the qubit state -- hence perfect conversion efficiency. This is nothing but the Quantum Zeno effect. For larger angle $\theta$ though, the energy conversion efficiency is smaller, and measurement behaves as a partially ordered energy source.

\subsection{Elevator}
\begin{figure}

\begin{minipage}[l]{0.45\textwidth} 
\includegraphics[scale=.7]{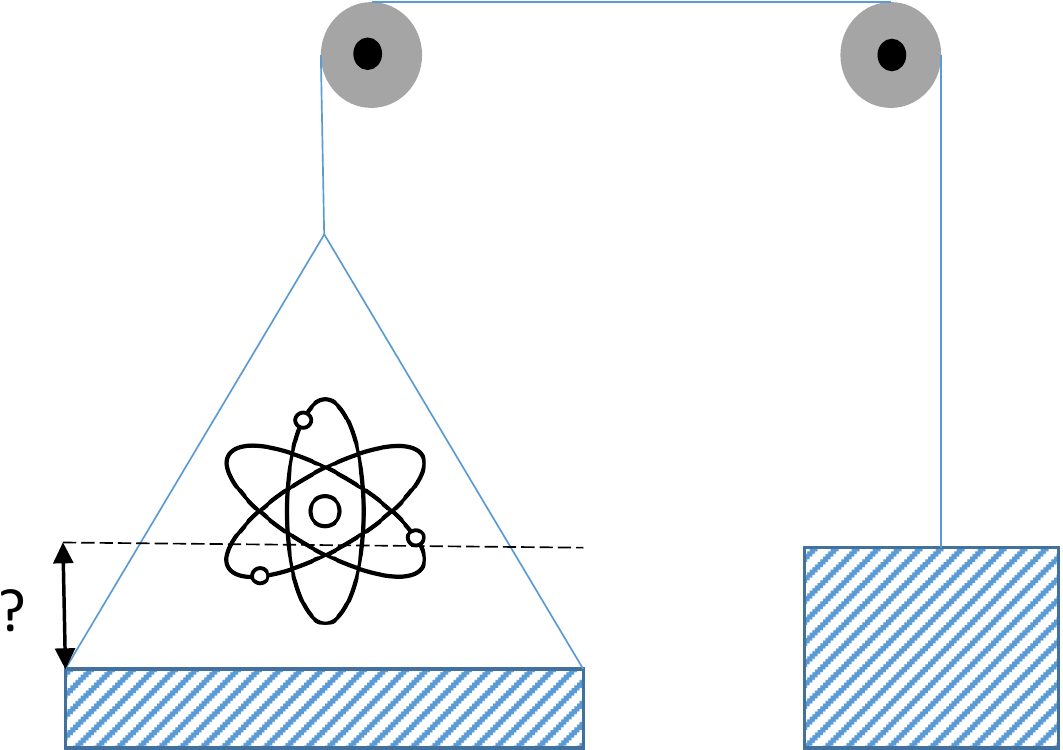}
\end{minipage}\hfill
\begin{minipage}[l]{0.05\textwidth} 
\end{minipage}\hfill
\begin{minipage}[l]{0.50\textwidth} 
\vspace{0.8cm}
\caption{An elevator platform is counterbalanced with an equal weight suspended by pulleys.  An atom is placed on the elevator platform, with the goal of lifting it with no work done by the counterbalanced weight.  Making generalized quantum measurements of the position of the atom, together with lifting the platform by the shown amount if the atom is found not to be near the platform can accomplish this.}
\label{fig:elevator}
\end{minipage}

\end{figure}
A system with continuous degrees of system was considered in Ref.~\cite{Elouard18}.  There, a particle in a gravitational potential is supported by the floor, which is movable with a pulley and counterweight (i.e. an elevator), shown in Fig.~\ref{fig:elevator}.  Unlike the two-level system described in the previous section, the measurement made is also on a continuous degree of freedom.  A sharp position measurement of the particle would not commute with the Hamiltonian, but turns out to require an infinite amount of energy.  We therefore can considered a coarse-grained measurement, and ask if the particle is within some window $\varepsilon$ of the floor, or not.  Even this type of measurement is too costly energetically, so it is advantageous to soften the measurement and put in a grey region.  We want a measurement that will tell us {\it for sure} the particle is not within some region within $\varepsilon$ of the floor, but the other possibilities have a grey region of width $w-\varepsilon$, where the particle might be here or it might be there.  One such choice was proposed in Ref.~\cite{Elouard18} which is to adopt the Kraus operator for one output to be
\be
M_0(x) = \begin{cases}
0, & 0 < x <\varepsilon \\
\sin \frac{\pi (x-\varepsilon)}{2(w-\varepsilon)} ,&  \varepsilon < x < w \\
1, & \varepsilon > w.
\end{cases}
\label{krauss}
\ee
and another measurement output corresponds to the Kraus operator $M_1(x) = \sqrt{1-M_0^2}$.  The Kraus (measurement) operators are defined by the nature of the coupling between system in meter, and the manner in which the meter is measured.

Starting the system in its ground state corresponds to an extended wavefunction in position space, with a characteristic size of $x_0 = (\hbar^2 / 2 m F)^{1/3}$, where $F$ is the constant force pushing on the particle. By making $\varepsilon$ to be of the order of this characteristic size $x_0$, the modified quantum state is very nearly equal to the ground state of a modified geometry, where the elevator floor is advanced by an amount $\varepsilon$.  Therefore, by having the controller simply lift the atom by an amount $\varepsilon$ when outcome ``0'' is registered by the controller, it is certain that the atom is not going to be disturbed by the movement of the elevator floor (because it has zero wavefunction there), and it is certain that by stopping the elevator at a position of $\varepsilon$, the new state of the particle is nearly the new ground state of the advanced potential, so little energy is wasted in the relaxation step of settling back to the ground state before the next step in the engine cycle (this can be done by coupling to a zero temperature bath; the photon environment, for example, whereupon a spontaneous emission event will restore the particle to its ground state). Upon successful operation of the engine, the energy stored in the meter degree of freedom is transferred to the system, resulting in useful work being (stochastically) done to the system, such the elevator will lift the particle without consuming any net energy of the system.

\subsection{Atom and Piston Engine}
\begin{figure}
\includegraphics[scale=.7]{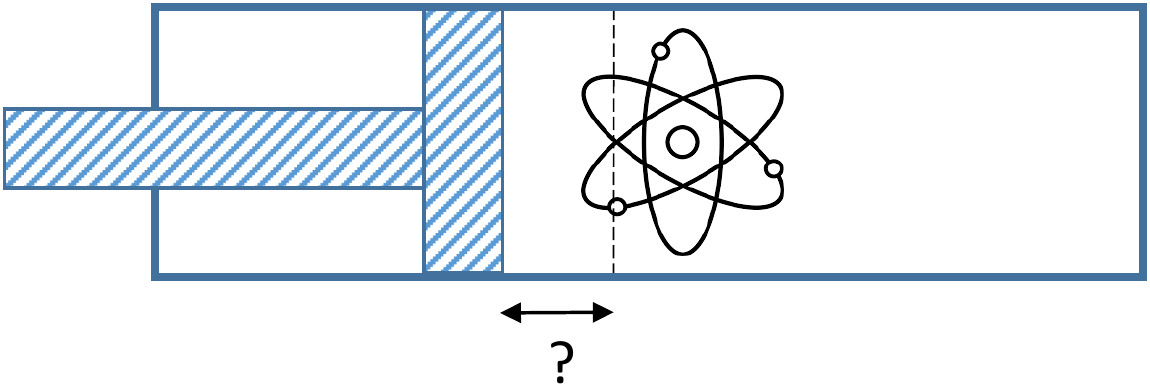}
\caption{An atom is placed inside of a cylindrical container with a movable piston.  An engine is created by making a generalized quantum measurement of the position of the atom, together with inserting the piston into the cylinder by the indicated amount if the atom is found not to be near the piston.  The engine cycle is completed by adiabatically expanding the piston and allowing the atom to do work on the external agent.}
\label{fig:piston}
\end{figure}
A somewhat simpler system we can work out explicitly is that of a particle in a box, where one side of the box is a move-able piston, shown in Fig.~\ref{fig:piston}.  As before, we prepare the system in its ground state, and carry out a generalized measurement of the form (\ref{krauss}).  If the particle is found to be not close to the piston, we can simply insert the piston quickly, with no work done.  Unlike the case of the elevator, to extract work from the system, we take the post-measurement state and adiabatically allow the particle to push the piston out, doing useful work.  When the piston is back in the original configuration, we then let the particle relax back to its ground state by exchanging energy with a zero temperature bath.

The system is quite simple to analyze quantitatively.  
The ground state of a particle in a box is simply
\be
\psi_g(x) = \sqrt{\frac{2}{L}}\sin (\pi x/L),
\ee
where $L$ is the length of the one-dimensional box, with an energy 
\be
E_g = \frac{\hbar^2 \pi^2}{2 m L^2}.
\ee
We can engineer the appropriate measurement operators that create a perfect state reset to the ground state of the next engine stage.  As was noted in Ref.~\cite{Elouard18}, the ideal case of perfect energy to work conversion is when the post measurement state of the system is prepared in the ground state of the next engine cycle stage, and in the unsuccessful case, the state is left undisturbed.  We will see that while the first is possible here, the second is not.  For this example engine, the ideal new state corresponds to a ground state of the box, with the piston advanced by $\epsilon$, given by
\be
{\tilde \psi}_g(x) = \sqrt{\frac{2}{L-\epsilon}}\sin \left[\frac{\pi (x-\epsilon)}{L-\epsilon}\right],
\ee
\begin{figure}
\includegraphics[scale=.6]{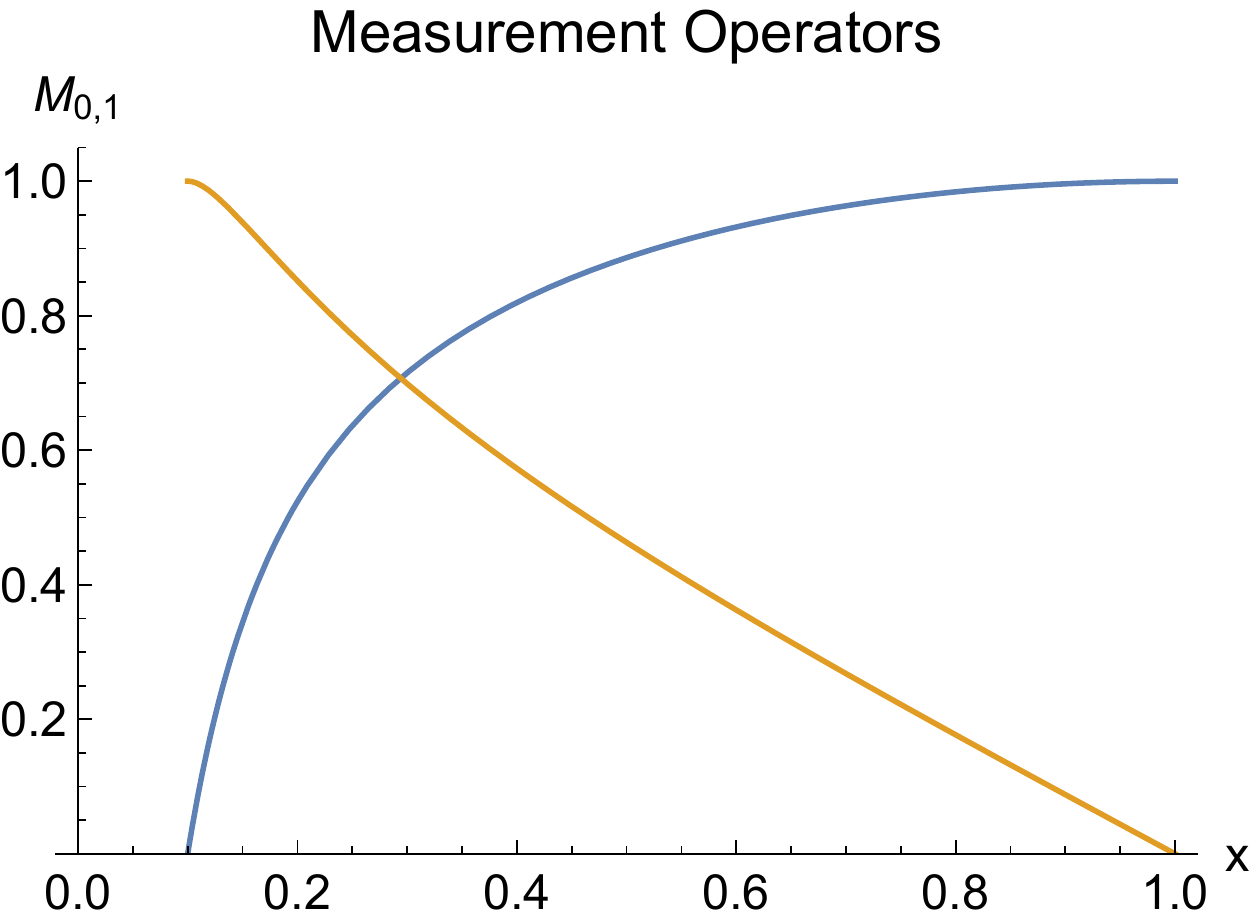}
\caption{The two measurement operators $M_0$ (blue) and $M_1$ (orange) are plotted versus position (in units of $L$) for the value $\varepsilon = 0.1 L$.  For positive positions less than $\epsilon$, the operators are continued to either 0 (for $M_0$) and 1 (for $M_1$). }
\label{fig:mos}
\end{figure}
where now $x \in [\epsilon, L]$, and its value is taken to be 0 from $x \in [0, \epsilon]$.  The new conditioned quantum state, given meter result $0$ or $1$ is given by $\psi_{0,1} \propto M_{0,1} \psi_g$, where $M_{0,1}$ are the new Kraus (measurement) operators.  This system is nice because an exact construction of perfectly efficient measurement operator is possible in the position basis, one of which is simply proportional to the ratio of the post and pre measurement state, and the other must obey the POVM condition, $M_0^2+ M_1^2 = 1$,
\be
M_0 =  \begin{cases}
0, & 0 < x <\varepsilon \\
\frac{(L-\varepsilon) \sin\left[\frac{\pi (x-\epsilon)}{L-\epsilon}\right]}{L \sin (\pi x/L)},&
 \varepsilon < x < L.
 \end{cases}
\qquad M_1 = \sqrt{1-M_0^2}.
\ee
These functions are plotted in Fig.~\ref{fig:mos} for the value $\epsilon = 0.1 L$.  They are analytic and well behaved functions except at the point $x=\epsilon$, however, this gives a negligible energetic cost because the discontinuity in $M_1$ is only in the second derivative, and the first derivative discontinuity in $M_0$ has a zero value of the function.

It is interesting to find the expected work gained from the measurement process.  The energy gain $\Delta E$ from a successful advance of the piston is simply the difference of the ground state energies,
\be
\Delta E = \frac{\pi^2 \hbar^2}{2m L^2} \left[\left(1-\frac{\varepsilon}{L}\right)^{-2} - 1\right].\label{gain}
\ee
Therefore the average work extracted is this energy gain (\ref{gain}) times the probability of a ``0'' meter event.  This may be calculated exactly,
\be
P_0 = \int_{\epsilon}^L dx M_0^2 \psi_g^2 = \left (1-\frac{\varepsilon}{L}\right)^3.
\ee
We see that as $\varepsilon/L \rightarrow 0$, the probability quickly limits to 1; that is, the particle is nearly always far away from the region $x \in (0, \varepsilon)$.  However, the probability decays only as a power law; even if the measurement is made with a left/right measurement (i.e. $\varepsilon = L/2$), then we still have a 1/8 chance of getting a successful measurement outcome.
\begin{figure}
\includegraphics[scale=.6]{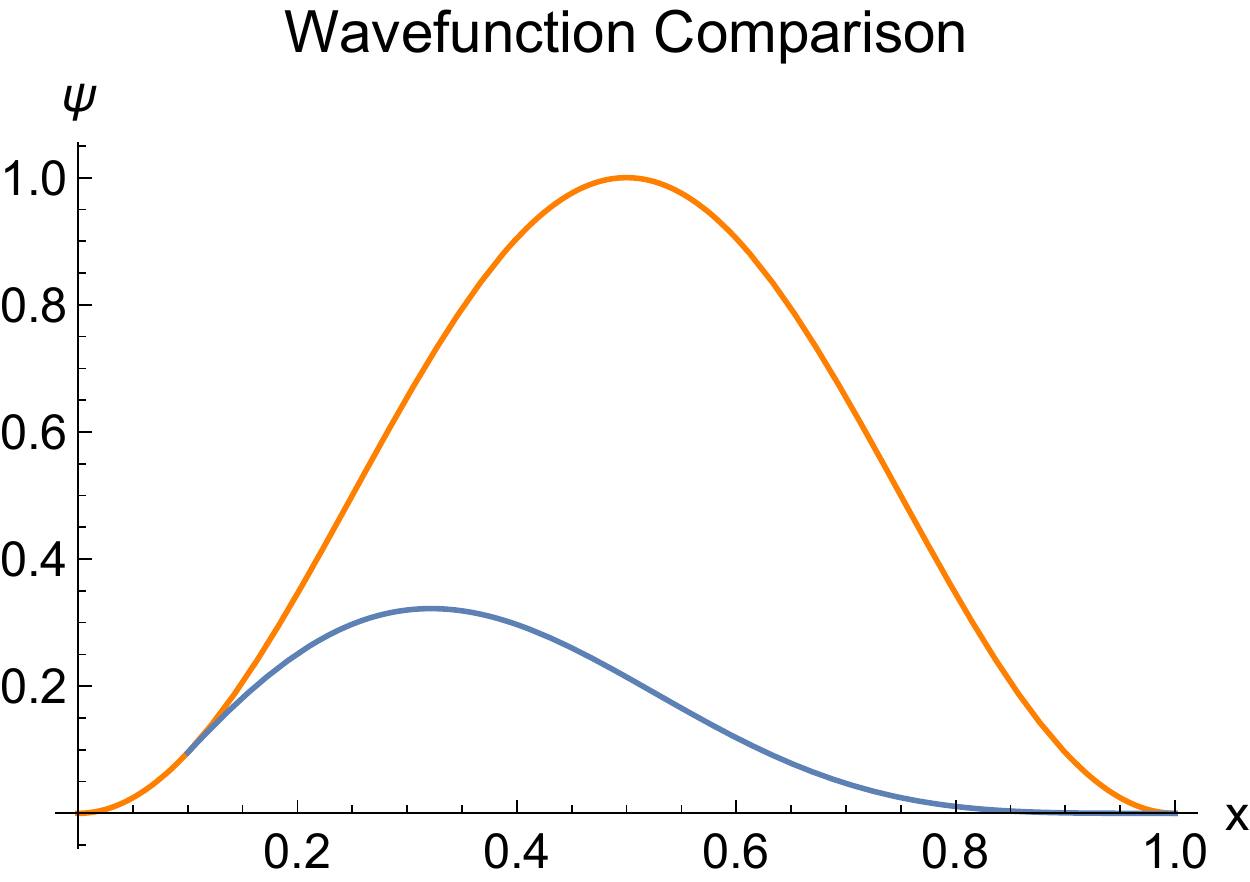}
\caption{The unnormalized wavefunctions $\psi_g(x)$ (orange) and $M_1\psi_g$ (blue) are plotted versus position in units of $L$ for the choice $\varepsilon =0.1 L$.  For $x>\varepsilon$, the wavefunctions coincide, but for $x > \varepsilon$, they quickly diverge, indicating that the disturbed wavefunction has an expected energy of much more than the ground state, resulting in wasted energy when this outcome occurs.}
\label{fig:comparepsi}
\end{figure}

By construction, the post-measurement state is the new ground state of the the potential when the piston is moved over a distance $\varepsilon$.  Consequently, we can extract all of that energy as useful work by adiabatically expanding the piston to the original position.  We now concern ourselves with the other case, when the meter returns the outcome ``1'', indicating that the particle is close to the piston, so we cannot move it without doing work on the particle.  In this case, the disturbed wavefunction is plotted in Fig.~\ref{fig:comparepsi} and compared with the original ground state (both are unnormalized).  We see that by construction, for $x<\varepsilon$, the wavefunctions coincide, but that for $x>\varepsilon$, they differ quite a lot, which will result in a much higher energy from the ground state.  This example then realizes an interesting situation that the successful events produce all the work and none of the wasted energy, whereas the failed events produce none of the work, and all of the wasted energy.

\begin{figure}
    \centering
    \begin{minipage}{0.4\textwidth}
        \centering
        \includegraphics[width=\textwidth]{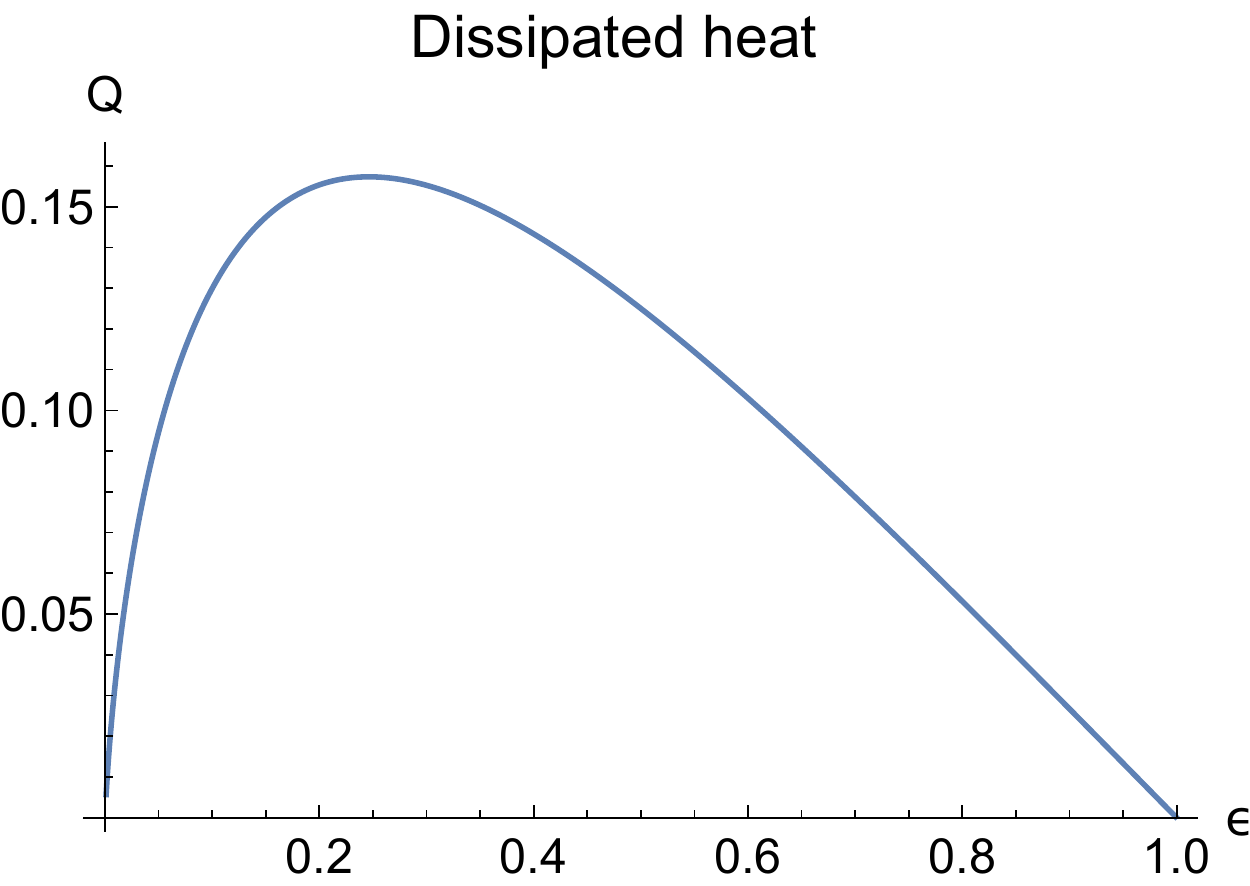} 
        \caption{The dissipated heat is plotted in units of $E_g$ versus $\varepsilon$ in units of $L$.}
          \label{qheat}
    \end{minipage}\hfill
    \begin{minipage}{0.4\textwidth}
        \centering
        \includegraphics[width=\textwidth]{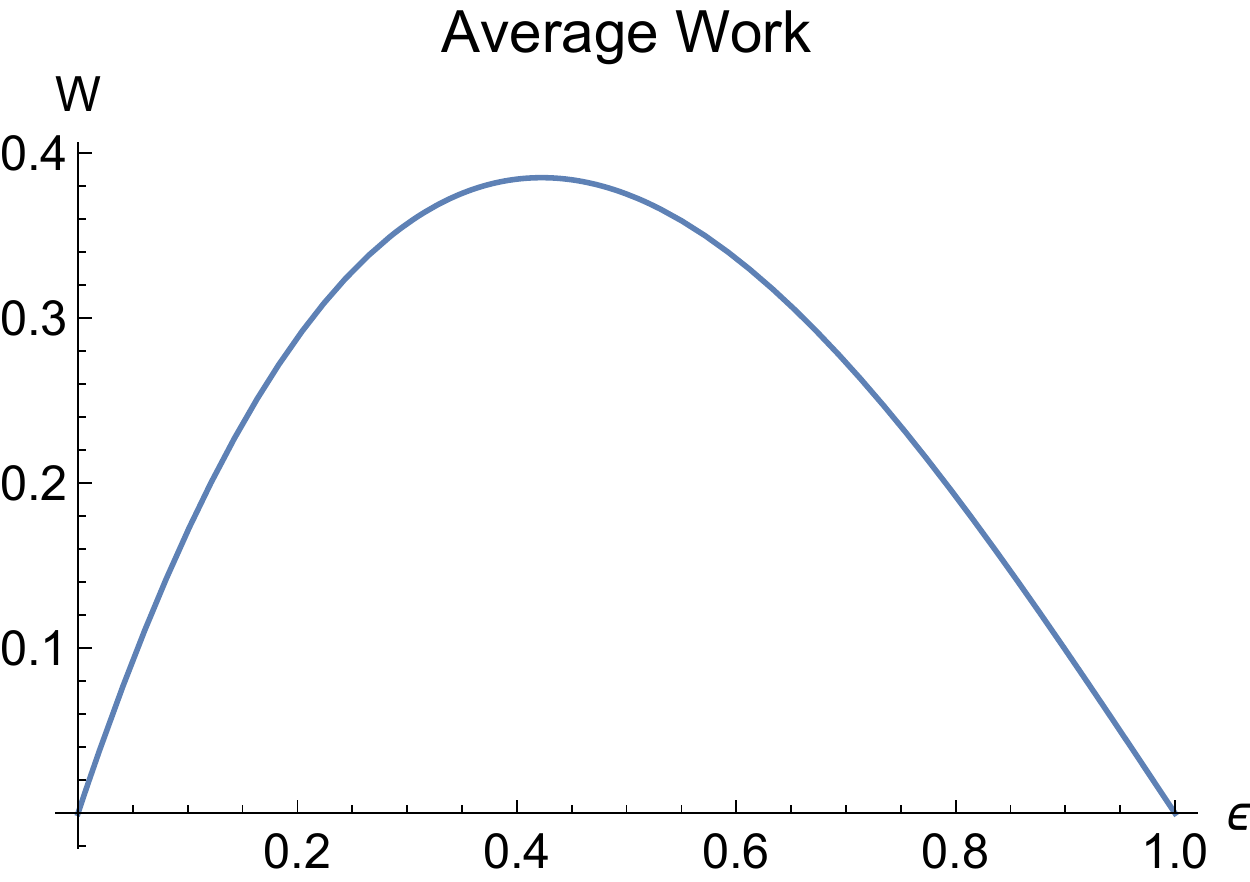} 
        \caption{The average work is plotted in units of $E_g$ versus $\varepsilon$ in units of $L$.}
         \label{work}
    \end{minipage}\hfill
       \begin{minipage}{0.4\textwidth}
        \centering
        \includegraphics[width=\textwidth]{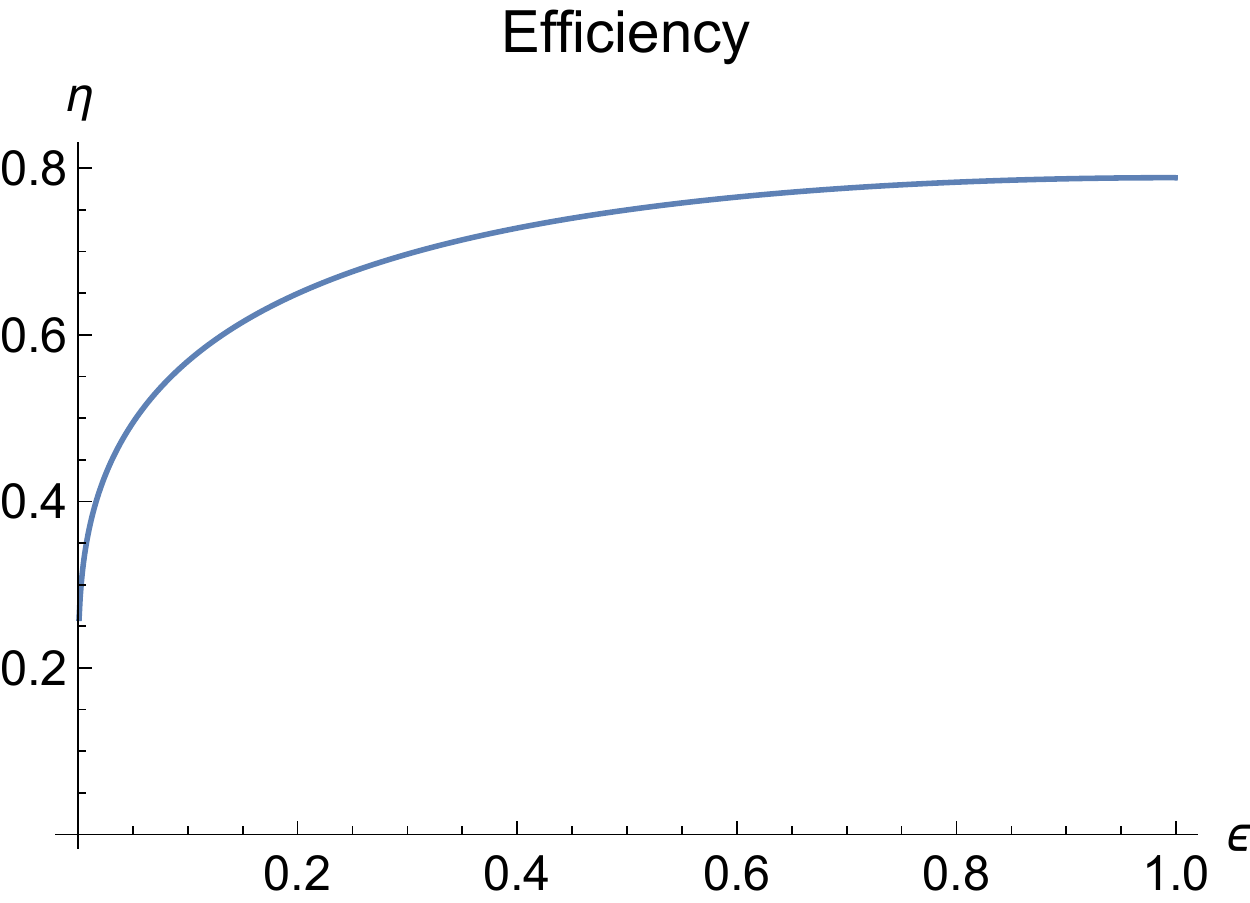} 
        \caption{The engine efficiency is plotted in units of $E_g$ versus $\varepsilon$ in units of $L$.}
         \label{efficiency}
    \end{minipage}
\end{figure}

The total added energy the measurement gives to the system is, on average,
\be
{\cal E}_q = \sum_{\alpha = 0,1} P_\alpha \la \phi_\alpha | H | \psi_\alpha \ra - E_g,
\ee
where $\phi_\alpha = M_\alpha |\psi_g\ra /||M_\alpha |\psi_g\ra||$ are the normalized, post-measurement states.  This energy can be divided into average work, $W$ and the energy wasted under the form of dissipated heat $Q$.  From the analysis above, the average work is given by
\be
W = E_g \left(1 - \frac{\varepsilon}{L}\right) \left( 1 - \left(1 - \frac{\varepsilon}{L}\right)^2\right),
\ee
and the dissipated heat by
\be
Q = \la M_1 \psi_g | H | M_1 \psi_g\ra - E_g P_1,
\ee
The expectation value of the energy is calculated in position space by breaking the integral into two parts, $x\in (0, \varepsilon)$ and $x \in (\varepsilon, L)$, the second of which is carried out numerically.  We plot $Q$ in Fig.~\ref{qheat}, $W$ in Fig.~\ref{work}, and the efficiency of the process, defined as 
\be
\eta = \frac{W}{Q+W},
\ee
in Fig.~\ref{efficiency}.  Here we have neglected the memory reset cost for simplicity, considering the ideal case of a zero temperature measuring device. In contrast with the simple qubit engine or optimized elevator, the conversion of the energy received by the qubit into work is not perfect, resulting in an efficiency below $1$ even when neglecting memory reset cost.

\section{Unread measurements as a hot bath: measurement engine without feedback}

Another strategy that has been explored \cite{Yi17,Ding18} is to design engine which are powered by an unread measurement process. If the outcome of the measurement is not read, the state of the system being measured is prepared in a mixture of the different possible output states, which in the case where the measured observable does not commute with the system's Hamiltonian, contains more energy than the initial state. In this case, the measurement can be seen as an effective environment that prepares a non-thermal mixed state. Such an engineered environment can be used to replace the hot thermal reservoir in a heat engine cycle. 

The original proposals in Refs~\cite{Yi17,Ding18} involve 4-stroke engines of the following form.

\begin{enumerate}
\item {\it Adiabatic compression}: Starting in thermal equilibirium at temperature T, a parameter of the Hamiltonian is varied adiabatically in a direction increasing the energy. This step costs work $W_1>0$.
\item {\it Measurement}: A measurement is performed in a basis different from the energy eigenbasis. The oscillator receives energy ${\cal E}_M>0$.
\item {\it Adiabatic expansion}: The frequency is brought back to its initial value, allowing for work extraction, denoted $W_2 < 0$.
\item {\it Thermalization}: The state is brought back to thermal state, dissipating heat denoted $Q<0$.
\end{enumerate}

This protocol can cover a lot of different situations, due to the freedom in choosing the unitary transformation involved in the compression and expansion steps, and the measurement basis. This protocol is not limited to projective measurement neither, such that the strength of the measurement provides one more knob to turn. 

In the case of a qubit of initial Hamiltonian $H_i = \hbar\omega_i\sigma_z$, a natural protocol for the compression/expansion steps is to vary the frequency from $\omega_i$ to $\omega_f>\omega_i$ (compression) and back (expansion). Then the optimum measurement is in a basis in the equator of the Bloch sphere, e.g. the eigenstates $\ket{\pm} = (1/\sqrt 2)(\ket{e}\pm\ket{g})$ of $\sigma_x$. This simple case, considered in Ref.~\cite{Yi17} turns out to be exactly equivalent in terms of performances to an Otto engine, i.e. the same protocol but involving thermalization with an actual thermal bath at temperature $T_H > T$ instead of the measurement, in the limit $T_H \gg \omega_{i,f}$. This equivalence can be immediately deduced from the fact that the unread measurement of $\sigma_x$ yields the qubit in the maximally mixed state $\rho_\text{mm} = \idop/2$, which is also the thermal equilibrium state in the infinite temperature limit. Note that even though the Carnot efficiency is 1 when using an infinite temperature hot bath, the present protocol, just as the usual Otto engine, does not allow to reach unit efficiency but instead $1-\omega_i/\omega_f$.

This strict equivalence between the measurement and heat bath is specific to the qubit case, and to the chosen unitary for the work extraction which does not involve coherences, but only changes the level spacing. In Ref~\cite{Yi17}, the authors analyze the case of a particle in a potential and measured weakly in the position basis. In this case, the action of the measurement cannot be reduced to that of a thermal bath, but is characterized by the transition rates between the particle energy eigenstates it generates. These rates have an impact on the power extracted from the engine, even though the efficiency turns out to match that of an Otto engine once again.

\begin{figure}
  \begin{minipage}[l]{0.4\textwidth}
  \includegraphics[width=\textwidth]{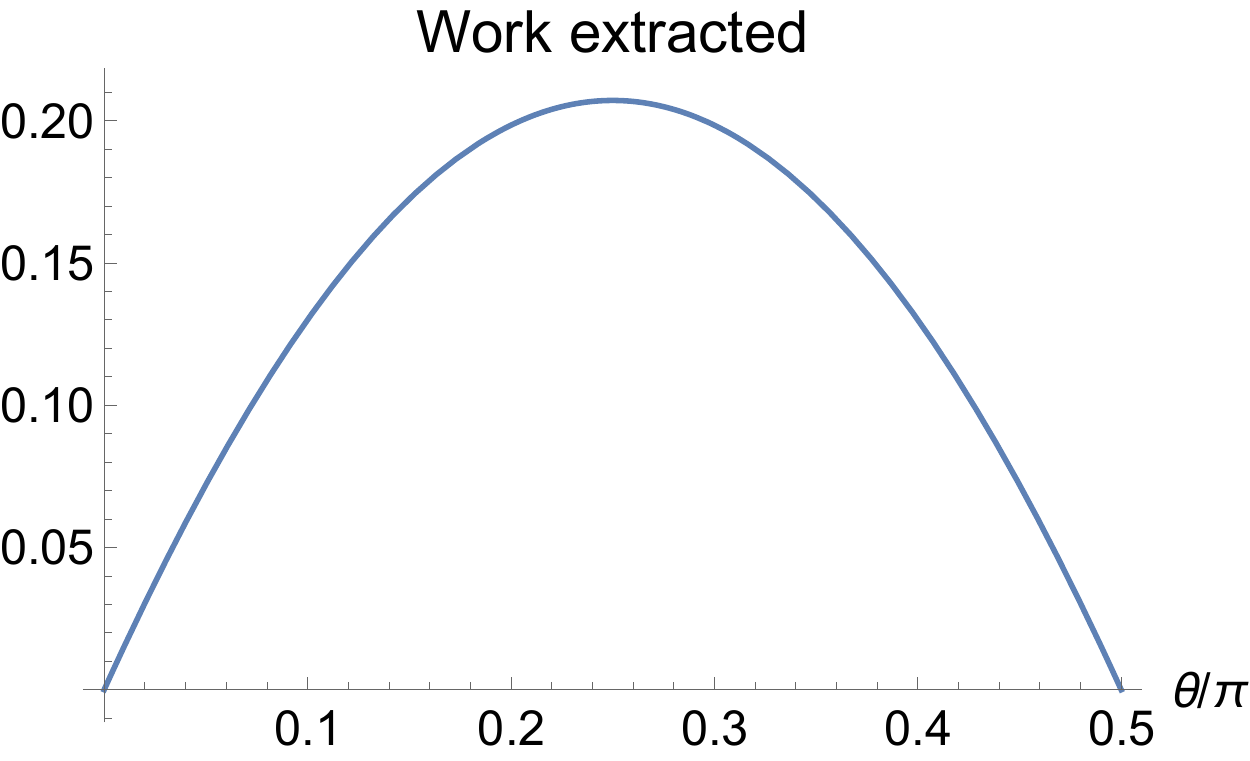}\\
   \includegraphics[width=\textwidth]{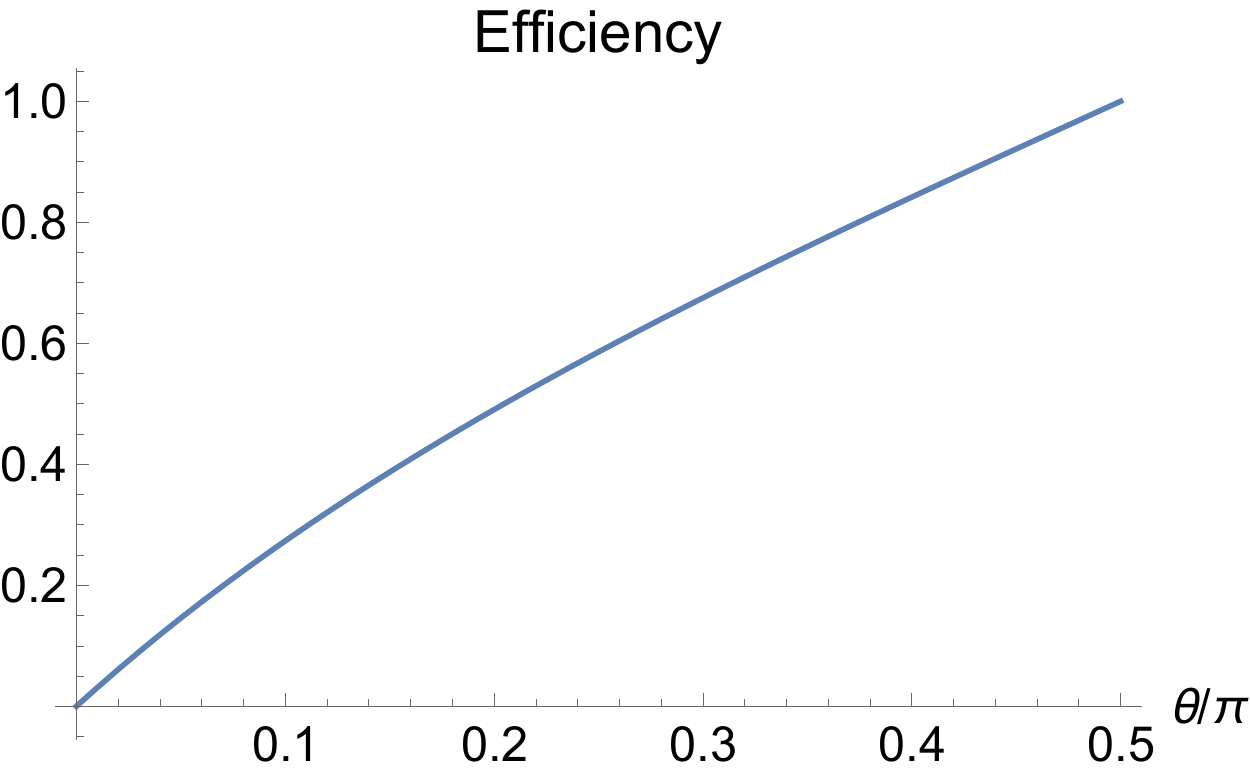}
  \caption{Work extracted per cycle (top) and efficiency (bottom) of the 4-stroke qubit engine powered by the unread $\sigma_x$ measurement against the angle $\theta$ of the first coherent rotation.}
         \label{f:4strokesPerf} 
    \end{minipage}\hfill
    \begin{minipage}[r]{0.4\textwidth} 
   \vspace{1.5cm}
 \includegraphics[width=\textwidth]{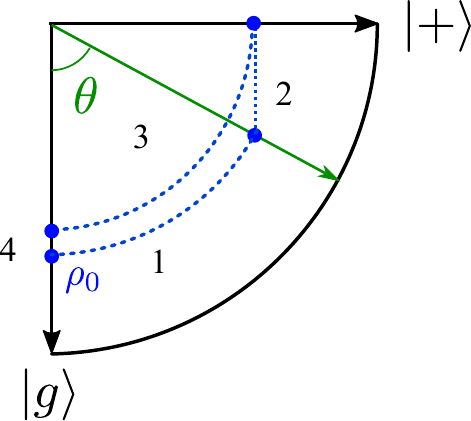}
    \vspace{0.2cm}
  \caption{Cycle in the Bloch sphere of the 4-stroke qubit engine: (1) coherent rotation of angle $\theta$, (2) unread measurement of $\sigma_x$, (3) coherent rotation of angle $-\pi/2$, (4) thermalization.}
     \label{f:4strokes}
  \end{minipage}\hfill
\end{figure}

In order to exploit more the differences between the measurement process and a thermal reservoir, and recover some of the nice features of measurement-driven engines presented earlier, e.g. the possibility of unit efficiency, we go back to the protocol above and use a unitary that involve coherences in the energy eigenbasis. In the case of a qubit, we replace the compression and the expansion with a coherent rotation around the y axis of the Bloch sphere as in the first engine of Section \ref{sec:qubit}. We keep the measurement to be in the $\{\ket{+},\ket{-}\}$ basis. The state of the qubit at the beginning of the cycle is $\rho_0 = (1+z_\text{th}\sigma_z)/2$, with $z_\text{th}$ defined as the thermal value of the z-coordinate. The coherent rotation brings it to $\rho_1 = (1+z_\text{th}\sigma_\theta)/2$, where $\sigma_\theta = \cos(\theta)\sigma_z + \sin(\theta)\sigma)x$, and costs a work $W_1 = -z_\text{th}\hbar\omega_0 \sin^2(\theta/2)>0$. Then, the measurement prepares $\rho_2 = (1+z\sin(\theta)\sigma_x)/2$, bringing energy ${\cal E}_M = - z_\text{th}\hbar\omega_0\cos(\theta)/2$. In step 3, a second coherent rotation of angle $-\pi/2$ brings the qubit back to a state diagonal in the $\{\ket{e},\ket{g}\}$ basis, allowing for work extraction such that $W_2 = z_\text{th}\hbar\omega_0\sin(\theta)/2 <0$. Finally, the last thermalization closes the cycle preparing $\rho_0$ and exchanging heat $Q = z_\text{th}\hbar\omega_0(1-\sin(\theta)) < 0$. The work output and the efficiency $\eta = - (W_1+W_2)/{\cal E}_M$ are plotted in Fig.~\ref{f:4strokesPerf} as a function of $\theta$, showing that the limit $\theta \to \pi/2$ allows to reach unit efficiency. Looking at the cycle in the Bloch sphere (see Fig.~\ref{f:4strokes}) allows us to see that this protocol is reminiscent of the one presented in Section \ref{sec:theory}, where the necessity of feedback is replaced by a thermalization process.

\begin{figure}
    \centering
    \begin{minipage}{0.4\textwidth}
        \centering
        \includegraphics[width=\textwidth]{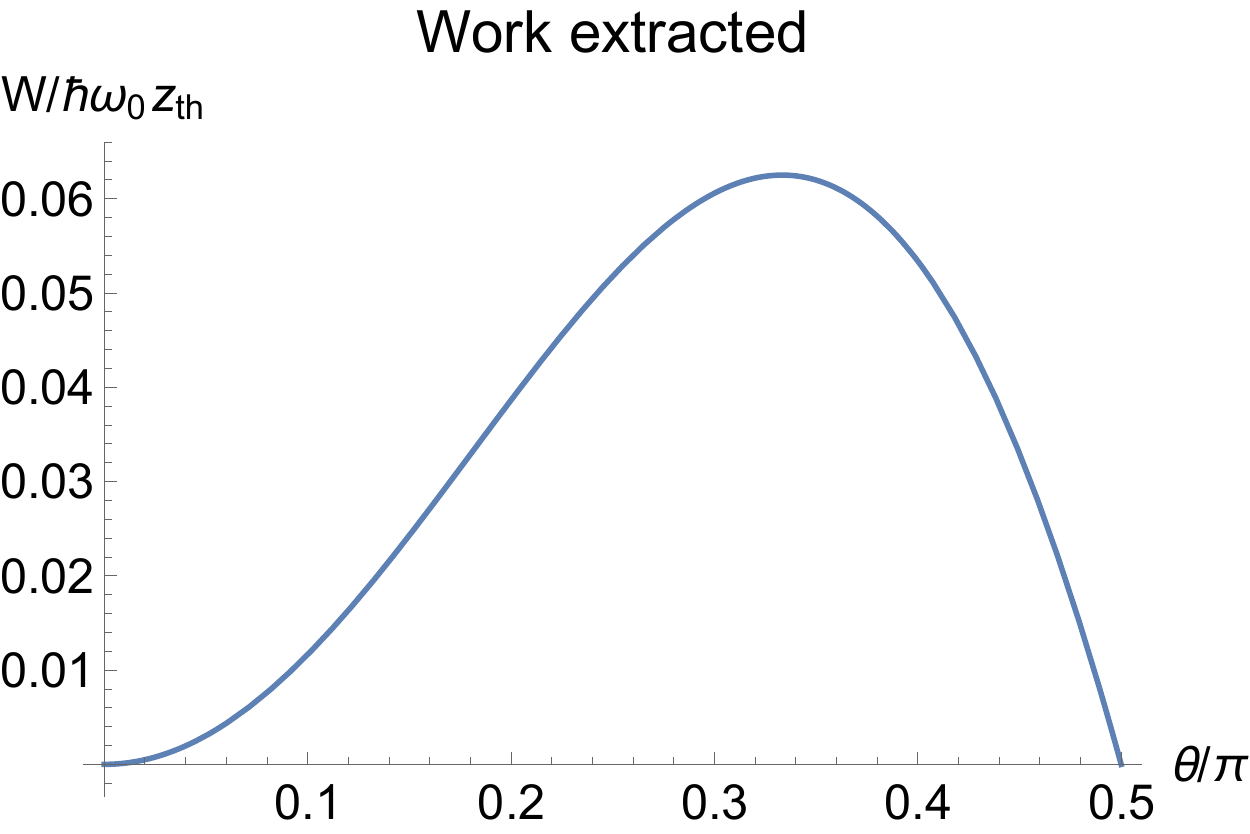} 
        \caption{Work extracted during the 3-stroke qubit engine as a function of the measurement basis angle $\theta$.}
          \label{f:3strokes_W}
    \end{minipage}\hfill
    \begin{minipage}{0.4\textwidth}
        \centering
        \includegraphics[width=\textwidth]{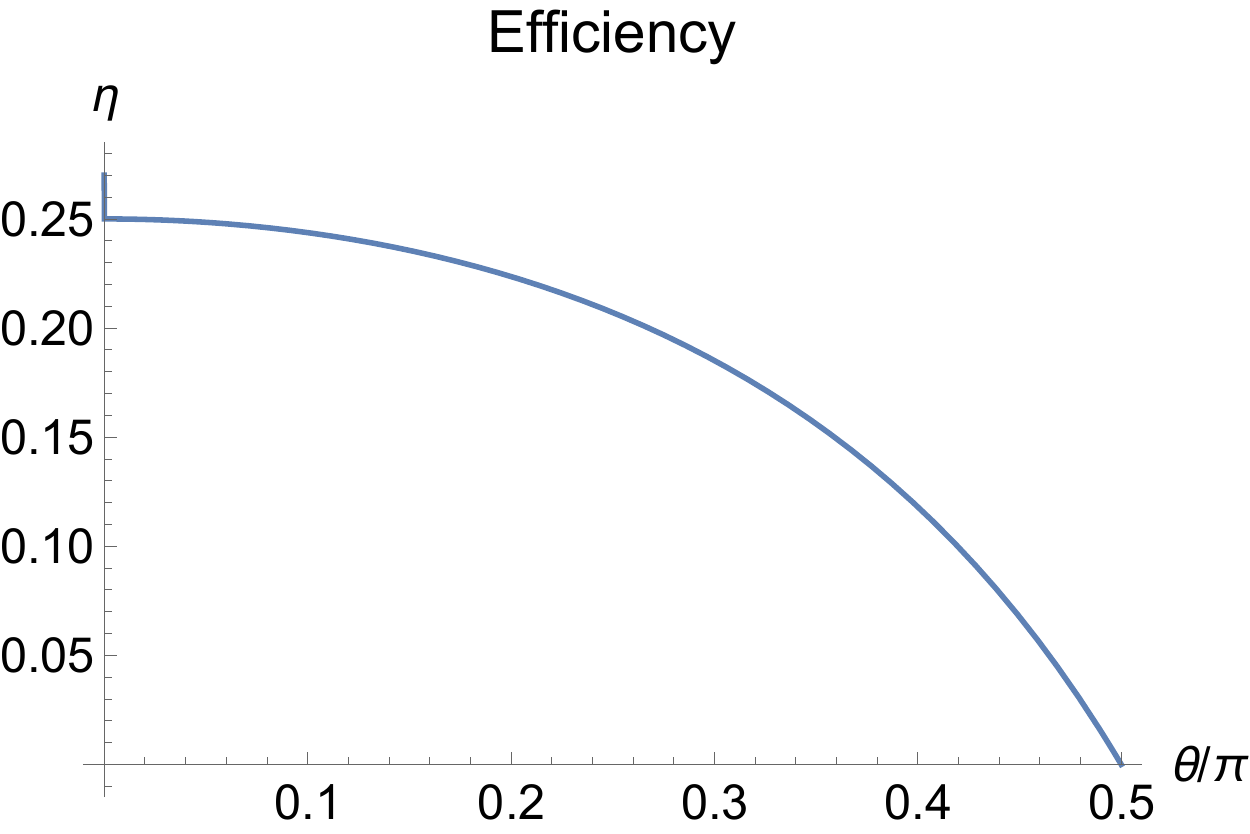} 
        \caption{Efficiency of the 3-stroke qubit engine as a function of the measurement basis angle $\theta$.}
         \label{f:3strokes_eta}
    \end{minipage}
\end{figure}

As a last comment, we note in contrast with a hot thermal reservoir, the measurement can be used to implement a 3-stroke engine. This is because the unread measurement process is able to transform directly a thermal equilibrium state into a state suitable for work extraction, i.e. which is not ``passive''\cite{Pusz78}. In contrast, any thermal state no matter how high the temperature, is passive, meaning that no unitary transformation can decrease its energy and therefore lead to average work extraction. As an illustration, we can once again consider the qubit, and apply a measurement directly on $\rho_0 = (1+z_\text{th}\sigma_z)/2$. This time, we chose the measurement basis to be $\{\ket{+_\phi},\ket{-_\phi}\}$, where $\ket{+_\phi} = \cos(\phi/2)\ket{e}+\sin(\phi/2)\ket{g}$ and  $\ket{-_\phi} = \sin(\phi/2)\ket{e}-\cos(\phi/2)\ket{g}$. Such measurement when not read prepares $\rho_1 = (1+z_\text{th}\cos(\phi)\sigma_\phi)/2$ providing energy ${\cal E}_M = -\hbar\omega_0 z_\text{th}\sin^2(\phi)$. A second stroke corresponds to work extraction via a coherent rotation preparing $\rho_2 = (1+z_\text{th}\cos(\phi)\sigma_z)/2$, allowing for work extraction $W = \hbar\omega_0 z_\text{th} \cos(\phi)(1-\cos(\phi)) <0$. Finally, the cycle is closed as before by a thermalization dissipating heat $Q = -{\cal E}_M - W <0$. In contrast with the 4-stroke engine, the measurement in the eigenbasis of $\sigma_x$ (i.e. $\phi = \pi/2$) is not interesting, as it prepares the infinite temperature thermal state, which is passive. The best efficiency, $1/4$, is obtained for $\phi\to 0$, and the maximum work output is obtained for $\phi = \pi/3$ (see Fig.~\ref{f:3strokes_W}-\ref{f:3strokes_eta}). Note that variations of this protocol can be imagined, e.g. replacing the coherent rotation for work extraction with the optimum process for work extraction from coherences introduced in Ref.~\cite{Kammerlander16}.

\section{Philosophy and Quantum Foundations}
Extending the notions of heat, work and entropy production to the quantum regime is an essential motivation of quantum thermodynamics. In the present case, there is still no consensus about the status of the energy provided by the quantum measurement channel. In the present paper, this energy appears as a resource for quantum engines, such that the measuring channel plays a role similar to a bath - justifying the denomination of ``quantum heat". Conversely, measurement induced energetic fluctuations can also be classified as non-conservative work by other approaches that rather focused on the energy necessary to perform a measurement \cite{Sagawa08}. 

We believe that the current debates about this status simply reflect the variety of interpretations about the measurement postulate and the subsequent opinions on the nature of quantum randomness \cite{RS}. Quantum measurement is seen as a fundamental source of randomness, noise and irreversibility in some approaches, and a practical way to extract information on the system in others. In the future, it will be an essential task to investigate whether these conceptual differences bring two visions of the world that can be physically discriminated, or if they rather pertain to metaphysical debates. If the demarcation between heat and work in the quantum regime is metaphysical, it should solve the current debates. In the opposite case, it could allow to suggest new experiments based on quantum thermodynamics to discriminate between the different interpretations of quantum mechanics. The Holy Grail would be to identify an experiment that can discriminate whether ``quantum heat” is heat or work such as the Bell test can discriminate whether two systems can be described by local hidden variables or not.

\section{Conclusions}
We have discussed several previously proposed quantum measurement based engines, as well as introduced several new ones.  These engines are based on purely quantum resources, and have direct connection to classical thermal heat engines, although there are specific disparities.  The atom and piston engine we discussed had the interesting feature that because we engineered the measurement such that the successful outcome produced the ground state of the system when the piston is advanced by the proper amount, the successful events account for all the work, but none of the wasted energy.  On the other hand, the failed event, when the atom is found to be too near the piston to move produces none of the work, but all of the quantum heat.  The result is an engine that is not perfectly efficient, but can still reach around 0.8 efficiency.  

We also introduced a four-stroke qubit measurement engine that is unconditional.  Different from previous proposals, the work extraction process does not require knowing the measurement outcome, but involves the coherences of the qubit:  the qubit has coherent rotations rather than simply adiabatically changing the level splitting.  This engine can attain perfect efficiency.  We also introduced a variant three-stroke engine that could not be accomplished by simply replacing the measurement process with a hot reservoir.

We have also pointed out that the way different researchers think about the quantum measurement, whether it is seen as a fundamental source of randomness versus simply a way to extract information from the system can be either reconciled by different metaphysical interpretations of the same quantum heat, or put in scientific conflict by implying different physical consequences.  Even if the former is true, we hope these new quantum engines can winnow down the possible interpretations and help disentangle the ontological from the epistemological.

\begin{acknowledgements}
We thank the John Templeton Foundation for sponsoring the ``Quantum  Limits  of  Knowledge” meeting in Copenhagen, Denmark at the Niels Bohr Institute, as well as Eugene Polzik for organizing it and inviting us to present our research.  ANJ and CE's work was supported by the U.S.
Department of Energy (DOE), Office of Science, Basic Energy Sciences (BES) under Award No. DE-SC-467 0017890.
We thank Joe Eberly for helpful discussions.
\end{acknowledgements}

\bibliography{refs}

\begin{thebibliography}{32}%
\makeatletter
\providecommand \@ifxundefined [1]{%
 \@ifx{#1\undefined}
}%
\providecommand \@ifnum [1]{%
 \ifnum #1\expandafter \@firstoftwo
 \else \expandafter \@secondoftwo
 \fi
}%
\providecommand \@ifx [1]{%
 \ifx #1\expandafter \@firstoftwo
 \else \expandafter \@secondoftwo
 \fi
}%
\providecommand \natexlab [1]{#1}%
\providecommand \enquote  [1]{``#1''}%
\providecommand \bibnamefont  [1]{#1}%
\providecommand \bibfnamefont [1]{#1}%
\providecommand \citenamefont [1]{#1}%
\providecommand \href@noop [0]{\@secondoftwo}%
\providecommand \href [0]{\begingroup \@sanitize@url \@href}%
\providecommand \@href[1]{\@@startlink{#1}\@@href}%
\providecommand \@@href[1]{\endgroup#1\@@endlink}%
\providecommand \@sanitize@url [0]{\catcode `\\12\catcode `\$12\catcode
  `\&12\catcode `\#12\catcode `\^12\catcode `\_12\catcode `\%12\relax}%
\providecommand \@@startlink[1]{}%
\providecommand \@@endlink[0]{}%
\providecommand \url  [0]{\begingroup\@sanitize@url \@url }%
\providecommand \@url [1]{\endgroup\@href {#1}{\urlprefix }}%
\providecommand \urlprefix  [0]{URL }%
\providecommand \Eprint [0]{\href }%
\providecommand \doibase [0]{http://dx.doi.org/}%
\providecommand \selectlanguage [0]{\@gobble}%
\providecommand \bibinfo  [0]{\@secondoftwo}%
\providecommand \bibfield  [0]{\@secondoftwo}%
\providecommand \translation [1]{[#1]}%
\providecommand \BibitemOpen [0]{}%
\providecommand \bibitemStop [0]{}%
\providecommand \bibitemNoStop [0]{.\EOS\space}%
\providecommand \EOS [0]{\spacefactor3000\relax}%
\providecommand \BibitemShut  [1]{\csname bibitem#1\endcsname}%
\let\auto@bib@innerbib\@empty
\bibitem [{\citenamefont {Bohr}(1928)}]{Bohr28}%
  \BibitemOpen
  \bibfield  {author} {\bibinfo {author} {\bibfnamefont {N.}~\bibnamefont
  {Bohr}},\ }\bibfield  {title} {\enquote {\bibinfo {title} {{The Quantum
  Postulate and the Recent Development of Atomic Theory1}},}\ }\href {\doibase
  10.1038/121580a0} {\bibfield  {journal} {\bibinfo  {journal} {Nature}\
  }\textbf {\bibinfo {volume} {121}},\ \bibinfo {pages} {580--590} (\bibinfo
  {year} {1928})}\BibitemShut {NoStop}%
\bibitem [{\citenamefont {Wiseman}\ and\ \citenamefont
  {Milburn}(2009)}]{WisemanBook}%
  \BibitemOpen
  \bibfield  {author} {\bibinfo {author} {\bibfnamefont {Howard~M.}\
  \bibnamefont {Wiseman}}\ and\ \bibinfo {author} {\bibfnamefont {Gerard~J.}\
  \bibnamefont {Milburn}},\ }\href {\doibase 10.1017/CBO9780511813948} {\emph
  {\bibinfo {title} {Quantum Measurement and Control}}}\ (\bibinfo  {publisher}
  {Cambridge University Press},\ \bibinfo {year} {2009})\BibitemShut {NoStop}%
\bibitem [{\citenamefont {Nogues}\ \emph {et~al.}(1999)\citenamefont {Nogues},
  \citenamefont {Rauschenbeutel}, \citenamefont {Osnaghi}, \citenamefont
  {Brune}, \citenamefont {Raimond},\ and\ \citenamefont {Haroche}}]{Nogues99}%
  \BibitemOpen
  \bibfield  {author} {\bibinfo {author} {\bibfnamefont {G.}~\bibnamefont
  {Nogues}}, \bibinfo {author} {\bibfnamefont {A.}~\bibnamefont
  {Rauschenbeutel}}, \bibinfo {author} {\bibfnamefont {S.}~\bibnamefont
  {Osnaghi}}, \bibinfo {author} {\bibfnamefont {M.}~\bibnamefont {Brune}},
  \bibinfo {author} {\bibfnamefont {J.~M.}\ \bibnamefont {Raimond}}, \ and\
  \bibinfo {author} {\bibfnamefont {S.}~\bibnamefont {Haroche}},\ }\bibfield
  {title} {\enquote {\bibinfo {title} {{Seeing a single photon without
  destroying it}},}\ }\href {\doibase 10.1038/22275} {\bibfield  {journal}
  {\bibinfo  {journal} {Nature}\ }\textbf {\bibinfo {volume} {400}},\ \bibinfo
  {pages} {239--242} (\bibinfo {year} {1999})}\BibitemShut {NoStop}%
\bibitem [{\citenamefont {Korotkov}\ and\ \citenamefont
  {Jordan}(2006)}]{korotkov2006undoing}%
  \BibitemOpen
  \bibfield  {author} {\bibinfo {author} {\bibfnamefont {Alexander~N}\
  \bibnamefont {Korotkov}}\ and\ \bibinfo {author} {\bibfnamefont {Andrew~N}\
  \bibnamefont {Jordan}},\ }\bibfield  {title} {\enquote {\bibinfo {title}
  {Undoing a weak quantum measurement of a solid-state qubit},}\ }\href@noop {}
  {\bibfield  {journal} {\bibinfo  {journal} {Physical review letters}\
  }\textbf {\bibinfo {volume} {97}},\ \bibinfo {pages} {166805} (\bibinfo
  {year} {2006})}\BibitemShut {NoStop}%
\bibitem [{\citenamefont {Katz}\ \emph {et~al.}(2008)\citenamefont {Katz},
  \citenamefont {Neeley}, \citenamefont {Ansmann}, \citenamefont {Bialczak},
  \citenamefont {Hofheinz}, \citenamefont {Lucero}, \citenamefont
  {O’Connell}, \citenamefont {Wang}, \citenamefont {Cleland}, \citenamefont
  {Martinis} \emph {et~al.}}]{katz2008reversal}%
  \BibitemOpen
  \bibfield  {author} {\bibinfo {author} {\bibfnamefont {Nadav}\ \bibnamefont
  {Katz}}, \bibinfo {author} {\bibfnamefont {Matthew}\ \bibnamefont {Neeley}},
  \bibinfo {author} {\bibfnamefont {M}~\bibnamefont {Ansmann}}, \bibinfo
  {author} {\bibfnamefont {Radoslaw~C}\ \bibnamefont {Bialczak}}, \bibinfo
  {author} {\bibfnamefont {M}~\bibnamefont {Hofheinz}}, \bibinfo {author}
  {\bibfnamefont {Erik}\ \bibnamefont {Lucero}}, \bibinfo {author}
  {\bibfnamefont {A}~\bibnamefont {O’Connell}}, \bibinfo {author}
  {\bibfnamefont {H}~\bibnamefont {Wang}}, \bibinfo {author} {\bibfnamefont
  {AN}~\bibnamefont {Cleland}}, \bibinfo {author} {\bibfnamefont {John~M}\
  \bibnamefont {Martinis}},  \emph {et~al.},\ }\bibfield  {title} {\enquote
  {\bibinfo {title} {Reversal of the weak measurement of a quantum state in a
  superconducting phase qubit},}\ }\href@noop {} {\bibfield  {journal}
  {\bibinfo  {journal} {Physical review letters}\ }\textbf {\bibinfo {volume}
  {101}},\ \bibinfo {pages} {200401} (\bibinfo {year} {2008})}\BibitemShut
  {NoStop}%
\bibitem [{\citenamefont {Williams}\ and\ \citenamefont
  {Jordan}(2008)}]{williams2008weak}%
  \BibitemOpen
  \bibfield  {author} {\bibinfo {author} {\bibfnamefont {Nathan~S}\
  \bibnamefont {Williams}}\ and\ \bibinfo {author} {\bibfnamefont {Andrew~N}\
  \bibnamefont {Jordan}},\ }\bibfield  {title} {\enquote {\bibinfo {title}
  {Weak values and the leggett-garg inequality in solid-state qubits},}\
  }\href@noop {} {\bibfield  {journal} {\bibinfo  {journal} {Physical review
  letters}\ }\textbf {\bibinfo {volume} {100}},\ \bibinfo {pages} {026804}
  (\bibinfo {year} {2008})}\BibitemShut {NoStop}%
\bibitem [{\citenamefont {Dixon}\ \emph {et~al.}(2009)\citenamefont {Dixon},
  \citenamefont {Starling}, \citenamefont {Jordan},\ and\ \citenamefont
  {Howell}}]{dixon2009ultrasensitive}%
  \BibitemOpen
  \bibfield  {author} {\bibinfo {author} {\bibfnamefont {P~Ben}\ \bibnamefont
  {Dixon}}, \bibinfo {author} {\bibfnamefont {David~J}\ \bibnamefont
  {Starling}}, \bibinfo {author} {\bibfnamefont {Andrew~N}\ \bibnamefont
  {Jordan}}, \ and\ \bibinfo {author} {\bibfnamefont {John~C}\ \bibnamefont
  {Howell}},\ }\bibfield  {title} {\enquote {\bibinfo {title} {Ultrasensitive
  beam deflection measurement via interferometric weak value amplification},}\
  }\href@noop {} {\bibfield  {journal} {\bibinfo  {journal} {Physical review
  letters}\ }\textbf {\bibinfo {volume} {102}},\ \bibinfo {pages} {173601}
  (\bibinfo {year} {2009})}\BibitemShut {NoStop}%
\bibitem [{\citenamefont {Goggin}\ \emph {et~al.}(2011)\citenamefont {Goggin},
  \citenamefont {Almeida}, \citenamefont {Barbieri}, \citenamefont {Lanyon},
  \citenamefont {O’brien}, \citenamefont {White},\ and\ \citenamefont
  {Pryde}}]{goggin2011violation}%
  \BibitemOpen
  \bibfield  {author} {\bibinfo {author} {\bibfnamefont {ME}~\bibnamefont
  {Goggin}}, \bibinfo {author} {\bibfnamefont {MP}~\bibnamefont {Almeida}},
  \bibinfo {author} {\bibfnamefont {Marco}\ \bibnamefont {Barbieri}}, \bibinfo
  {author} {\bibfnamefont {BP}~\bibnamefont {Lanyon}}, \bibinfo {author}
  {\bibfnamefont {JL}~\bibnamefont {O’brien}}, \bibinfo {author}
  {\bibfnamefont {AG}~\bibnamefont {White}}, \ and\ \bibinfo {author}
  {\bibfnamefont {GJ}~\bibnamefont {Pryde}},\ }\bibfield  {title} {\enquote
  {\bibinfo {title} {Violation of the leggett--garg inequality with weak
  measurements of photons},}\ }\href@noop {} {\bibfield  {journal} {\bibinfo
  {journal} {Proceedings of the National Academy of Sciences}\ }\textbf
  {\bibinfo {volume} {108}},\ \bibinfo {pages} {1256--1261} (\bibinfo {year}
  {2011})}\BibitemShut {NoStop}%
\bibitem [{\citenamefont {Kocsis}\ \emph {et~al.}(2011)\citenamefont {Kocsis},
  \citenamefont {Braverman}, \citenamefont {Ravets}, \citenamefont {Stevens},
  \citenamefont {Mirin}, \citenamefont {Shalm},\ and\ \citenamefont
  {Steinberg}}]{Kocsis11}%
  \BibitemOpen
  \bibfield  {author} {\bibinfo {author} {\bibfnamefont {Sacha}\ \bibnamefont
  {Kocsis}}, \bibinfo {author} {\bibfnamefont {Boris}\ \bibnamefont
  {Braverman}}, \bibinfo {author} {\bibfnamefont {Sylvain}\ \bibnamefont
  {Ravets}}, \bibinfo {author} {\bibfnamefont {Martin~J.}\ \bibnamefont
  {Stevens}}, \bibinfo {author} {\bibfnamefont {Richard~P.}\ \bibnamefont
  {Mirin}}, \bibinfo {author} {\bibfnamefont {L.~Krister}\ \bibnamefont
  {Shalm}}, \ and\ \bibinfo {author} {\bibfnamefont {Aephraim~M.}\ \bibnamefont
  {Steinberg}},\ }\bibfield  {title} {\enquote {\bibinfo {title} {{Observing
  the Average Trajectories of Single Photons in a Two-Slit Interferometer}},}\
  }\href {\doibase 10.1126/science.1202218} {\bibfield  {journal} {\bibinfo
  {journal} {Science}\ }\textbf {\bibinfo {volume} {332}},\ \bibinfo {pages}
  {1170--1173} (\bibinfo {year} {2011})}\BibitemShut {NoStop}%
\bibitem [{\citenamefont {Chantasri}\ \emph {et~al.}(2013)\citenamefont
  {Chantasri}, \citenamefont {Dressel},\ and\ \citenamefont
  {Jordan}}]{chantasri2013action}%
  \BibitemOpen
  \bibfield  {author} {\bibinfo {author} {\bibfnamefont {A}~\bibnamefont
  {Chantasri}}, \bibinfo {author} {\bibfnamefont {Justin}\ \bibnamefont
  {Dressel}}, \ and\ \bibinfo {author} {\bibfnamefont {Andrew~N}\ \bibnamefont
  {Jordan}},\ }\bibfield  {title} {\enquote {\bibinfo {title} {Action principle
  for continuous quantum measurement},}\ }\href@noop {} {\bibfield  {journal}
  {\bibinfo  {journal} {Physical Review A}\ }\textbf {\bibinfo {volume} {88}},\
  \bibinfo {pages} {042110} (\bibinfo {year} {2013})}\BibitemShut {NoStop}%
\bibitem [{\citenamefont {Murch}\ \emph {et~al.}(2013)\citenamefont {Murch},
  \citenamefont {Weber}, \citenamefont {Macklin},\ and\ \citenamefont
  {Siddiqi}}]{Murch13}%
  \BibitemOpen
  \bibfield  {author} {\bibinfo {author} {\bibfnamefont {K.~W.}\ \bibnamefont
  {Murch}}, \bibinfo {author} {\bibfnamefont {S.~J.}\ \bibnamefont {Weber}},
  \bibinfo {author} {\bibfnamefont {C.}~\bibnamefont {Macklin}}, \ and\
  \bibinfo {author} {\bibfnamefont {I.}~\bibnamefont {Siddiqi}},\ }\bibfield
  {title} {\enquote {\bibinfo {title} {{Observing single quantum trajectories
  of a superconducting quantum bit}},}\ }\href {\doibase 10.1038/nature12539}
  {\bibfield  {journal} {\bibinfo  {journal} {nature}\ }\textbf {\bibinfo
  {volume} {502}},\ \bibinfo {pages} {211--214} (\bibinfo {year}
  {2013})}\BibitemShut {NoStop}%
\bibitem [{\citenamefont {Jordan}(2013)}]{jordan2013quantum}%
  \BibitemOpen
  \bibfield  {author} {\bibinfo {author} {\bibfnamefont {Andrew~N}\
  \bibnamefont {Jordan}},\ }\bibfield  {title} {\enquote {\bibinfo {title}
  {Quantum physics: Watching the wavefunction collapse},}\ }\href@noop {}
  {\bibfield  {journal} {\bibinfo  {journal} {Nature}\ }\textbf {\bibinfo
  {volume} {502}},\ \bibinfo {pages} {177} (\bibinfo {year}
  {2013})}\BibitemShut {NoStop}%
\bibitem [{\citenamefont {Weber}\ \emph {et~al.}(2014)\citenamefont {Weber},
  \citenamefont {Chantasri}, \citenamefont {Dressel}, \citenamefont {Jordan},
  \citenamefont {Murch},\ and\ \citenamefont {Siddiqi}}]{weber2014mapping}%
  \BibitemOpen
  \bibfield  {author} {\bibinfo {author} {\bibfnamefont {SJ}~\bibnamefont
  {Weber}}, \bibinfo {author} {\bibfnamefont {Areeya}\ \bibnamefont
  {Chantasri}}, \bibinfo {author} {\bibfnamefont {Justin}\ \bibnamefont
  {Dressel}}, \bibinfo {author} {\bibfnamefont {Andrew~N}\ \bibnamefont
  {Jordan}}, \bibinfo {author} {\bibfnamefont {KW}~\bibnamefont {Murch}}, \
  and\ \bibinfo {author} {\bibfnamefont {Irfan}\ \bibnamefont {Siddiqi}},\
  }\bibfield  {title} {\enquote {\bibinfo {title} {Mapping the optimal route
  between two quantum states},}\ }\href@noop {} {\bibfield  {journal} {\bibinfo
   {journal} {Nature}\ }\textbf {\bibinfo {volume} {511}},\ \bibinfo {pages}
  {570} (\bibinfo {year} {2014})}\BibitemShut {NoStop}%
\bibitem [{\citenamefont {Viza}\ \emph {et~al.}(2015)\citenamefont {Viza},
  \citenamefont {Mart{\'\i}nez-Rinc{\'o}n}, \citenamefont {Alves},
  \citenamefont {Jordan},\ and\ \citenamefont
  {Howell}}]{viza2015experimentally}%
  \BibitemOpen
  \bibfield  {author} {\bibinfo {author} {\bibfnamefont {Gerardo~I}\
  \bibnamefont {Viza}}, \bibinfo {author} {\bibfnamefont {Juli{\'a}n}\
  \bibnamefont {Mart{\'\i}nez-Rinc{\'o}n}}, \bibinfo {author} {\bibfnamefont
  {Gabriel~B}\ \bibnamefont {Alves}}, \bibinfo {author} {\bibfnamefont
  {Andrew~N}\ \bibnamefont {Jordan}}, \ and\ \bibinfo {author} {\bibfnamefont
  {John~C}\ \bibnamefont {Howell}},\ }\bibfield  {title} {\enquote {\bibinfo
  {title} {Experimentally quantifying the advantages of weak-value-based
  metrology},}\ }\href@noop {} {\bibfield  {journal} {\bibinfo  {journal}
  {Physical Review A}\ }\textbf {\bibinfo {volume} {92}},\ \bibinfo {pages}
  {032127} (\bibinfo {year} {2015})}\BibitemShut {NoStop}%
\bibitem [{\citenamefont {Chantasri}\ \emph {et~al.}(2016)\citenamefont
  {Chantasri}, \citenamefont {Kimchi-Schwartz}, \citenamefont {Roch},
  \citenamefont {Siddiqi},\ and\ \citenamefont
  {Jordan}}]{chantasri2016quantum}%
  \BibitemOpen
  \bibfield  {author} {\bibinfo {author} {\bibfnamefont {Areeya}\ \bibnamefont
  {Chantasri}}, \bibinfo {author} {\bibfnamefont {Mollie~E}\ \bibnamefont
  {Kimchi-Schwartz}}, \bibinfo {author} {\bibfnamefont {Nicolas}\ \bibnamefont
  {Roch}}, \bibinfo {author} {\bibfnamefont {Irfan}\ \bibnamefont {Siddiqi}}, \
  and\ \bibinfo {author} {\bibfnamefont {Andrew~N}\ \bibnamefont {Jordan}},\
  }\bibfield  {title} {\enquote {\bibinfo {title} {Quantum trajectories and
  their statistics for remotely entangled quantum bits},}\ }\href@noop {}
  {\bibfield  {journal} {\bibinfo  {journal} {Physical Review X}\ }\textbf
  {\bibinfo {volume} {6}},\ \bibinfo {pages} {041052} (\bibinfo {year}
  {2016})}\BibitemShut {NoStop}%
\bibitem [{\citenamefont {Naghiloo}\ \emph {et~al.}(2017)\citenamefont
  {Naghiloo}, \citenamefont {Tan}, \citenamefont {Harrington}, \citenamefont
  {Lewalle}, \citenamefont {Jordan},\ and\ \citenamefont
  {Murch}}]{naghiloo2017quantum}%
  \BibitemOpen
  \bibfield  {author} {\bibinfo {author} {\bibfnamefont {M}~\bibnamefont
  {Naghiloo}}, \bibinfo {author} {\bibfnamefont {D}~\bibnamefont {Tan}},
  \bibinfo {author} {\bibfnamefont {PM}~\bibnamefont {Harrington}}, \bibinfo
  {author} {\bibfnamefont {P}~\bibnamefont {Lewalle}}, \bibinfo {author}
  {\bibfnamefont {AN}~\bibnamefont {Jordan}}, \ and\ \bibinfo {author}
  {\bibfnamefont {KW}~\bibnamefont {Murch}},\ }\bibfield  {title} {\enquote
  {\bibinfo {title} {Quantum caustics in resonance-fluorescence
  trajectories},}\ }\href@noop {} {\bibfield  {journal} {\bibinfo  {journal}
  {Physical Review A}\ }\textbf {\bibinfo {volume} {96}},\ \bibinfo {pages}
  {053807} (\bibinfo {year} {2017})}\BibitemShut {NoStop}%
\bibitem [{\citenamefont {Chantasri}\ \emph {et~al.}(2018)\citenamefont
  {Chantasri}, \citenamefont {Atalaya}, \citenamefont {Hacohen-Gourgy},
  \citenamefont {Martin}, \citenamefont {Siddiqi},\ and\ \citenamefont
  {Jordan}}]{chantasri2018simultaneous}%
  \BibitemOpen
  \bibfield  {author} {\bibinfo {author} {\bibfnamefont {Areeya}\ \bibnamefont
  {Chantasri}}, \bibinfo {author} {\bibfnamefont {Juan}\ \bibnamefont
  {Atalaya}}, \bibinfo {author} {\bibfnamefont {Shay}\ \bibnamefont
  {Hacohen-Gourgy}}, \bibinfo {author} {\bibfnamefont {Leigh~S}\ \bibnamefont
  {Martin}}, \bibinfo {author} {\bibfnamefont {Irfan}\ \bibnamefont {Siddiqi}},
  \ and\ \bibinfo {author} {\bibfnamefont {Andrew~N}\ \bibnamefont {Jordan}},\
  }\bibfield  {title} {\enquote {\bibinfo {title} {Simultaneous continuous
  measurement of noncommuting observables: Quantum state correlations},}\
  }\href@noop {} {\bibfield  {journal} {\bibinfo  {journal} {Physical Review
  A}\ }\textbf {\bibinfo {volume} {97}},\ \bibinfo {pages} {012118} (\bibinfo
  {year} {2018})}\BibitemShut {NoStop}%
\bibitem [{\citenamefont {Dassonneville}\ \emph {et~al.}(2019)\citenamefont
  {Dassonneville}, \citenamefont {Ramos}, \citenamefont {Milchakov},
  \citenamefont {Planat}, \citenamefont {Dumur}, \citenamefont {Foroughi},
  \citenamefont {Puertas}, \citenamefont {Leger}, \citenamefont {Bharadwaj},
  \citenamefont {Delaforce}, \citenamefont {Naud}, \citenamefont
  {Hasch-Guichard}, \citenamefont
  {Garc{\ifmmode\acute{\imath}\else\'{\i}\fi}a-Ripoll}, \citenamefont {Roch},\
  and\ \citenamefont {Buisson}}]{Dassonneville19}%
  \BibitemOpen
  \bibfield  {author} {\bibinfo {author} {\bibfnamefont {R.}~\bibnamefont
  {Dassonneville}}, \bibinfo {author} {\bibfnamefont {T.}~\bibnamefont
  {Ramos}}, \bibinfo {author} {\bibfnamefont {V.}~\bibnamefont {Milchakov}},
  \bibinfo {author} {\bibfnamefont {L.}~\bibnamefont {Planat}}, \bibinfo
  {author} {\bibfnamefont {{\ifmmode\acute{E}\else\'{E}\fi}.}~\bibnamefont
  {Dumur}}, \bibinfo {author} {\bibfnamefont {F.}~\bibnamefont {Foroughi}},
  \bibinfo {author} {\bibfnamefont {J.}~\bibnamefont {Puertas}}, \bibinfo
  {author} {\bibfnamefont {S.}~\bibnamefont {Leger}}, \bibinfo {author}
  {\bibfnamefont {K.}~\bibnamefont {Bharadwaj}}, \bibinfo {author}
  {\bibfnamefont {J.}~\bibnamefont {Delaforce}}, \bibinfo {author}
  {\bibfnamefont {C.}~\bibnamefont {Naud}}, \bibinfo {author} {\bibfnamefont
  {W.}~\bibnamefont {Hasch-Guichard}}, \bibinfo {author} {\bibfnamefont
  {J.~J.}\ \bibnamefont {Garc{\ifmmode\acute{\imath}\else\'{\i}\fi}a-Ripoll}},
  \bibinfo {author} {\bibfnamefont {N.}~\bibnamefont {Roch}}, \ and\ \bibinfo
  {author} {\bibfnamefont {O.}~\bibnamefont {Buisson}},\ }\bibfield  {title}
  {\enquote {\bibinfo {title} {{Fast high fidelity quantum non-demolition qubit
  readout via a non-perturbative cross-Kerr coupling}},}\ }\href
  {https://arxiv.org/abs/1905.00271} {\bibfield  {journal} {\bibinfo  {journal}
  {arXiv}\ } (\bibinfo {year} {2019})},\ \Eprint
  {http://arxiv.org/abs/1905.00271} {1905.00271} \BibitemShut {NoStop}%
\bibitem [{\citenamefont {Rossi}\ \emph {et~al.}(2019)\citenamefont {Rossi},
  \citenamefont {Mason}, \citenamefont {Chen},\ and\ \citenamefont
  {Schliesser}}]{Rossi19}%
  \BibitemOpen
  \bibfield  {author} {\bibinfo {author} {\bibfnamefont {Massimiliano}\
  \bibnamefont {Rossi}}, \bibinfo {author} {\bibfnamefont {David}\ \bibnamefont
  {Mason}}, \bibinfo {author} {\bibfnamefont {Junxin}\ \bibnamefont {Chen}}, \
  and\ \bibinfo {author} {\bibfnamefont {Albert}\ \bibnamefont {Schliesser}},\
  }\bibfield  {title} {\enquote {\bibinfo {title} {{Observing and Verifying the
  Quantum Trajectory of a Mechanical Resonator}},}\ }\href {\doibase
  10.1103/PhysRevLett.123.163601} {\bibfield  {journal} {\bibinfo  {journal}
  {Phys. Rev. Lett.}\ }\textbf {\bibinfo {volume} {123}},\ \bibinfo {pages}
  {163601} (\bibinfo {year} {2019})}\BibitemShut {NoStop}%
\bibitem [{\citenamefont {Clausius}(1879)}]{Clausius1879}%
  \BibitemOpen
  \bibfield  {author} {\bibinfo {author} {\bibfnamefont {Rudolf}\ \bibnamefont
  {Clausius}},\ }\href@noop {} {\emph {\bibinfo {title} {{The Mechanical Theory
  of Heat}}}}\ (\bibinfo  {publisher} {Macmillan and Co.},\ \bibinfo {year}
  {1879})\BibitemShut {NoStop}%
\bibitem [{\citenamefont {Carnot}(1824)}]{Carnot1824}%
  \BibitemOpen
  \bibfield  {author} {\bibinfo {author} {\bibfnamefont {S.}~\bibnamefont
  {Carnot}},\ }\href@noop {} {\emph {\bibinfo {title}
  {{R{\ifmmode\acute{e}\else\'{e}\fi}flexions sur la puissance motrice du feu
  et sur les machines propres {\ifmmode\grave{a}\else\`{a}\fi}
  d{\ifmmode\acute{e}\else\'{e}\fi}velopper atte puissance}}}}\ (\bibinfo
  {publisher} {Bachelier Libraire},\ \bibinfo {year} {1824})\BibitemShut
  {NoStop}%
\bibitem [{\citenamefont {Elouard}\ \emph
  {et~al.}(2017{\natexlab{a}})\citenamefont {Elouard}, \citenamefont
  {Herrera-Mart{\ifmmode\acute{\imath}\else\'{\i}\fi}}, \citenamefont
  {Clusel},\ and\ \citenamefont
  {Auff{\ifmmode\grave{e}\else\`{e}\fi}ves}}]{Elouard17Role}%
  \BibitemOpen
  \bibfield  {author} {\bibinfo {author} {\bibfnamefont {Cyril}\ \bibnamefont
  {Elouard}}, \bibinfo {author} {\bibfnamefont {David~A.}\ \bibnamefont
  {Herrera-Mart{\ifmmode\acute{\imath}\else\'{\i}\fi}}}, \bibinfo {author}
  {\bibfnamefont {Maxime}\ \bibnamefont {Clusel}}, \ and\ \bibinfo {author}
  {\bibfnamefont {Alexia}\ \bibnamefont
  {Auff{\ifmmode\grave{e}\else\`{e}\fi}ves}},\ }\bibfield  {title} {\enquote
  {\bibinfo {title} {{The role of quantum measurement in stochastic
  thermodynamics}},}\ }\href {\doibase 10.1038/s41534-017-0008-4} {\bibfield
  {journal} {\bibinfo  {journal} {npj Quantum Inf.}\ }\textbf {\bibinfo
  {volume} {3}},\ \bibinfo {pages} {1--10} (\bibinfo {year}
  {2017}{\natexlab{a}})}\BibitemShut {NoStop}%
\bibitem [{\citenamefont {Elouard}\ \emph
  {et~al.}(2017{\natexlab{b}})\citenamefont {Elouard}, \citenamefont
  {Herrera-Mart{\ifmmode\acute{\imath}\else\'{\i}\fi}}, \citenamefont {Huard},\
  and\ \citenamefont {Auff{\ifmmode\grave{e}\else\`{e}\fi}ves}}]{Elouard17}%
  \BibitemOpen
  \bibfield  {author} {\bibinfo {author} {\bibfnamefont {Cyril}\ \bibnamefont
  {Elouard}}, \bibinfo {author} {\bibfnamefont {David}\ \bibnamefont
  {Herrera-Mart{\ifmmode\acute{\imath}\else\'{\i}\fi}}}, \bibinfo {author}
  {\bibfnamefont {Benjamin}\ \bibnamefont {Huard}}, \ and\ \bibinfo {author}
  {\bibfnamefont {Alexia}\ \bibnamefont
  {Auff{\ifmmode\grave{e}\else\`{e}\fi}ves}},\ }\bibfield  {title} {\enquote
  {\bibinfo {title} {{Extracting Work from Quantum Measurement in Maxwell's
  Demon Engines}},}\ }\href {\doibase 10.1103/PhysRevLett.118.260603}
  {\bibfield  {journal} {\bibinfo  {journal} {Phys. Rev. Lett.}\ }\textbf
  {\bibinfo {volume} {118}},\ \bibinfo {pages} {260603} (\bibinfo {year}
  {2017}{\natexlab{b}})}\BibitemShut {NoStop}%
\bibitem [{\citenamefont {Landauer}(1961)}]{landauer1961irreversibility}%
  \BibitemOpen
  \bibfield  {author} {\bibinfo {author} {\bibfnamefont {Rolf}\ \bibnamefont
  {Landauer}},\ }\bibfield  {title} {\enquote {\bibinfo {title}
  {Irreversibility and heat generation in the computing process},}\ }\href@noop
  {} {\bibfield  {journal} {\bibinfo  {journal} {IBM journal of research and
  development}\ }\textbf {\bibinfo {volume} {5}},\ \bibinfo {pages} {183--191}
  (\bibinfo {year} {1961})}\BibitemShut {NoStop}%
\bibitem [{\citenamefont {Bennett}(2003)}]{bennett2003notes}%
  \BibitemOpen
  \bibfield  {author} {\bibinfo {author} {\bibfnamefont {Charles~H}\
  \bibnamefont {Bennett}},\ }\bibfield  {title} {\enquote {\bibinfo {title}
  {Notes on landauer's principle, reversible computation, and maxwell's
  demon},}\ }\href@noop {} {\bibfield  {journal} {\bibinfo  {journal} {Studies
  In History and Philosophy of Science Part B: Studies In History and
  Philosophy of Modern Physics}\ }\textbf {\bibinfo {volume} {34}},\ \bibinfo
  {pages} {501--510} (\bibinfo {year} {2003})}\BibitemShut {NoStop}%
\bibitem [{\citenamefont {Elouard}\ and\ \citenamefont
  {Jordan}(2018)}]{Elouard18}%
  \BibitemOpen
  \bibfield  {author} {\bibinfo {author} {\bibfnamefont {Cyril}\ \bibnamefont
  {Elouard}}\ and\ \bibinfo {author} {\bibfnamefont {Andrew~N.}\ \bibnamefont
  {Jordan}},\ }\bibfield  {title} {\enquote {\bibinfo {title} {{Efficient
  Quantum Measurement Engines}},}\ }\href {\doibase
  10.1103/PhysRevLett.120.260601} {\bibfield  {journal} {\bibinfo  {journal}
  {Phys. Rev. Lett.}\ }\textbf {\bibinfo {volume} {120}},\ \bibinfo {pages}
  {260601} (\bibinfo {year} {2018})}\BibitemShut {NoStop}%
\bibitem [{\citenamefont {Yi}\ \emph {et~al.}(2017)\citenamefont {Yi},
  \citenamefont {Talkner},\ and\ \citenamefont {Kim}}]{Yi17}%
  \BibitemOpen
  \bibfield  {author} {\bibinfo {author} {\bibfnamefont {Juyeon}\ \bibnamefont
  {Yi}}, \bibinfo {author} {\bibfnamefont {Peter}\ \bibnamefont {Talkner}}, \
  and\ \bibinfo {author} {\bibfnamefont {Yong~Woon}\ \bibnamefont {Kim}},\
  }\bibfield  {title} {\enquote {\bibinfo {title} {{Single-temperature quantum
  engine without feedback control}},}\ }\href {\doibase
  10.1103/PhysRevE.96.022108} {\bibfield  {journal} {\bibinfo  {journal} {Phys.
  Rev. E}\ }\textbf {\bibinfo {volume} {96}},\ \bibinfo {pages} {022108}
  (\bibinfo {year} {2017})}\BibitemShut {NoStop}%
\bibitem [{\citenamefont {Ding}\ \emph {et~al.}(2018)\citenamefont {Ding},
  \citenamefont {Yi}, \citenamefont {Kim},\ and\ \citenamefont
  {Talkner}}]{Ding18}%
  \BibitemOpen
  \bibfield  {author} {\bibinfo {author} {\bibfnamefont {Xuehao}\ \bibnamefont
  {Ding}}, \bibinfo {author} {\bibfnamefont {Juyeon}\ \bibnamefont {Yi}},
  \bibinfo {author} {\bibfnamefont {Yong~Woon}\ \bibnamefont {Kim}}, \ and\
  \bibinfo {author} {\bibfnamefont {Peter}\ \bibnamefont {Talkner}},\
  }\bibfield  {title} {\enquote {\bibinfo {title} {{Measurement-driven single
  temperature engine}},}\ }\href {\doibase 10.1103/PhysRevE.98.042122}
  {\bibfield  {journal} {\bibinfo  {journal} {Phys. Rev. E}\ }\textbf {\bibinfo
  {volume} {98}},\ \bibinfo {pages} {042122} (\bibinfo {year}
  {2018})}\BibitemShut {NoStop}%
\bibitem [{\citenamefont {Pusz}\ and\ \citenamefont
  {Woronowicz}(1978)}]{Pusz78}%
  \BibitemOpen
  \bibfield  {author} {\bibinfo {author} {\bibfnamefont {W.}~\bibnamefont
  {Pusz}}\ and\ \bibinfo {author} {\bibfnamefont {S.~L.}\ \bibnamefont
  {Woronowicz}},\ }\bibfield  {title} {\enquote {\bibinfo {title} {{Passive
  states and KMS states for general quantum systems}},}\ }\href {\doibase
  10.1007/BF01614224} {\bibfield  {journal} {\bibinfo  {journal} {Commun. Math.
  Phys.}\ }\textbf {\bibinfo {volume} {58}},\ \bibinfo {pages} {273--290}
  (\bibinfo {year} {1978})}\BibitemShut {NoStop}%
\bibitem [{\citenamefont {Kammerlander}\ and\ \citenamefont
  {Anders}(2016)}]{Kammerlander16}%
  \BibitemOpen
  \bibfield  {author} {\bibinfo {author} {\bibfnamefont {P.}~\bibnamefont
  {Kammerlander}}\ and\ \bibinfo {author} {\bibfnamefont {J.}~\bibnamefont
  {Anders}},\ }\bibfield  {title} {\enquote {\bibinfo {title} {{Coherence and
  measurement in quantum thermodynamics}},}\ }\href {\doibase
  10.1038/srep22174} {\bibfield  {journal} {\bibinfo  {journal} {Sci. Rep.}\
  }\textbf {\bibinfo {volume} {6}},\ \bibinfo {pages} {22174} (\bibinfo {year}
  {2016})}\BibitemShut {NoStop}%
\bibitem [{\citenamefont {Sagawa}\ and\ \citenamefont {Ueda}(2008)}]{Sagawa08}%
  \BibitemOpen
  \bibfield  {author} {\bibinfo {author} {\bibfnamefont {Takahiro}\
  \bibnamefont {Sagawa}}\ and\ \bibinfo {author} {\bibfnamefont {Masahito}\
  \bibnamefont {Ueda}},\ }\bibfield  {title} {\enquote {\bibinfo {title}
  {{Minimal Energy Cost for Thermodynamic Information Processing: Measurement
  and Information Erasure}},}\ }\href {\doibase 10.1103/PhysRevLett.102.250602}
  {\bibfield  {journal} {\bibinfo  {journal} {arXiv}\ } (\bibinfo {year}
  {2008}),\ 10.1103/PhysRevLett.102.250602},\ \Eprint
  {http://arxiv.org/abs/0809.4098} {0809.4098} \BibitemShut {NoStop}%
\bibitem [{\citenamefont {Grangier}\ and\ \citenamefont
  {Auff\`eves}(2018)}]{RS}%
  \BibitemOpen
  \bibfield  {author} {\bibinfo {author} {\bibfnamefont {P.}~\bibnamefont
  {Grangier}}\ and\ \bibinfo {author} {\bibfnamefont {A.}~\bibnamefont
  {Auff\`eves}},\ }\bibfield  {title} {\enquote {\bibinfo {title} {{What is
  quantum in quantum randomness?}}}\ }\href {\doibase 10.1098/rsta.2017.0322}
  {\bibfield  {journal} {\bibinfo  {journal} {Philosophical Transactions of the
  Royal Society A: Mathematical, Physical and Engineering Sciences}\ }\textbf
  {\bibinfo {volume} {376}} (\bibinfo {year} {2018}),\
  10.1098/rsta.2017.0322}\BibitemShut {NoStop}%
\end{thebibliography}%

\end{document}